\documentclass[11pt]{article}

\usepackage{amsmath}
\usepackage{amssymb}
\usepackage{times}
\usepackage{graphicx}
\usepackage{color}
\usepackage{multirow}
\usepackage[dvipsnames]{xcolor}
\usepackage{natbib}
\usepackage{rotating}
\usepackage{bbm}
\usepackage{latexsym}
\usepackage{makecell}
\usepackage[dvipsnames]{xcolor}

\newcommand{\mbu}{\mathbf{u}}
\newcommand{\mbx}{\mathbf{x}}
\newcommand{\mby}{\mathbf{y}}
\newcommand{\R}{\mathbb{R}}

\begin{document}

\ \\
{\LARGE Sensitivity to control signals in triphasic rhythmic neural systems:  a comparative mechanistic analysis
\textit{via} infinitesimal local timing response curves}

\ \\
{\bf \large Zhuojun Yu}\\
zhuojun.yu@case.edu\\
Department of Mathematics, Applied Mathematics, and Statistics, Case Western Reserve University, Cleveland, OH 44106, USA\\
{\bf \large Jonathan E.~Rubin}\\
jonrubin@pitt.edu\\
Department of Mathematics, University of Pittsburgh, Pittsburgh, PA 15260, USA\\
{\bf \large Peter J.~Thomas}\\
pjthomas@case.edu\\
Department of Mathematics, Applied Mathematics, and Statistics, Department of Biology, Department of Electrical, Control and Systems Engineering, Case Western Reserve University, Cleveland, OH 44106, USA

Similar activity patterns may arise from model neural networks with distinct coupling properties and individual unit dynamics.
These similar patterns may, however, respond differently to parameter variations and, specifically, to tuning of inputs that represent control signals.
In this work, we analyze the responses resulting from modulation of a localized input in each of three classes of model neural networks that have been recognized in the literature for their capacity to produce robust three-phase rhythms: coupled fast-slow oscillators, near-heteroclinic oscillators, and threshold-linear networks.
Triphasic rhythms, in which each phase consists of a prolonged activation of a corresponding subgroup of neurons followed by a fast transition to another phase, represent a fundamental activity pattern observed across a range of central pattern generators underlying behaviors critical to survival, including respiration, locomotion, and feeding.
To perform our analysis, we extend the recently developed local timing response curve (lTRC), which allows us to characterize the timing effects due to perturbations, and we complement our lTRC approach with model-specific dynamical systems analysis. Interestingly, we observe disparate effects of similar perturbations across distinct model classes.
Thus, this work provides an  analytical framework for studying control of oscillations in nonlinear dynamical systems, and may help guide model selection in future efforts to study systems exhibiting triphasic rhythmic activity.



\section{Introduction}
\label{Introduction}

Many of the automatic behaviors fundamental to animal life, including respiration, digestion, and various forms of locomotion, represent multi-phase processes, in each of which a sequence of motor unit activations repeats rhythmically.
In each case, the behavior is driven by the generation of sustained, multi-phase rhythmic outputs by a corresponding neural circuit that, for automaticity, must be able to produce its characteristic activation pattern in the absence of temporally-varying inputs.
Nonetheless, inputs do serve important functions for these neuronal activity patterns; transient inputs may serve as triggers to turn rhythms and behaviors on and off, they may allow for interruption of rhythms (e.g., pausing of inspiration to allow for swallowing or vocalization) and switching between rhythms (e.g., transitions in swimming and walking in salamanders \citep{ijspeert2007swimming}), and they may serve as control signals to modulate the frequency or relative phase durations of ongoing rhythms.

The ubiquity of these rhythms, and the relative experimental accessibility of the central pattern generator (CPG) circuits that often produce them, have made them a common subject of scientific investigation. In the computational realm, past modeling works have presented several distinct mathematical frameworks that can generate repetitive, multi-phase rhythms.
In each case, the rhythms can exhibit extended activations of specific units in a circuit, occurring sequentially in each cycle.  Despite its functional importance, however, the topic of control of such circuits has received far less theoretical attention.
On the one hand, control of linear systems is a thoroughly developed topic (cf.~\cite{brockett2015finite}).
Fixed-point control for nonlinear systems is understood to the extent that one can approximate the nonlinear system under weak disturbances with a related linearized system (cf.~\cite{isidori1985nonlinear}).
Despite decades of development in these areas, however, control for nonlinear dynamical systems exhibiting limit cycle dynamics remains relative underdeveloped.
One reason for this lack of development may be that parametric control of limit cycles requires understanding the coordinated changes in both the shape and timing of trajectories upon adjustment of a control parameter.
For smooth systems, Floquet analysis can capture the effects of small perturbations, but in many naturally occurring control systems -- for instance in biological motor control systems -- there are switching surfaces or fast-slow dynamics that require more subtle treatment.
Such nonsmooth dynamics can occur at the point of making or breaking contact between the organism's body and an external substrate, or at hard boundaries for internal dynamics such as exclusion of negative firing rates in neural network CPG models formulated in the firing rate framework.

In this paper, we present a novel analysis of control\footnote{The term \emph{control} encompasses a broad set of technical problems.
In the context of mathematical physiology, it has long been understood that the regulation of e.g.~metabolic processes (such as breathing) is a primary example of ``control" in biological systems.  
As far back as 1954, Grodins et al.~asserted that ``The essence of physiology is regulation" \citep{grodins1954respiratory} and drew an explicit connection between control theory and the biology of respiratory control.  
In his seminal work on cybernetics, Wiener asserted a similar connection \citep{wiener1948cybernetics}.} in rhythm-generating neuronal circuits, achieved by applying and extending recently developed tools that allow \emph{linearized} (sometimes called \emph{variational}) analysis of the parametric variations induced in periodic orbits by changes in constant input levels.
As noted in previous work \citep{yakovenko2005,olypher2006,daun2009control,rubin2009multiple}, the functional control of CPGs requires not only the capacity to alter the overall output frequency but also, in general, a mechanism to tune the durations of specific phases within a rhythmic pattern, such as prolonging an inspiration without altering expiration or prolonging the stance phase of locomotion without altering the swing phase; in practice, however, changes in the duration of activity at one node in a coupled circuit may induce cascading changes in other nodes.
Therefore, in this paper we study how changing the input intensity to a single unit within a model CPG circuit impacts the durations of all phases within its rhythmic output cycle, with an eye towards the independent phase modulation property.

An important aspect of the work presented here is that we consider this control problem within several classes of mathematical models that naturally produce multi-phase oscillations and have appeared in the neuroscience literature.
One such modeling framework consists of relaxation oscillators coupled with mutually inhibitory synapses. The individual relaxation oscillator model includes dynamics on distinct fast and slow timescales and produces periodic dynamics featuring alternating, extended active and silent phases, with fast transitions between them.
This pattern provides a strong qualitative resemblance to the outputs of units within CPGs, and the fast-slow nature of the dynamics involved has proved convenient for a broad collection of analyses (cf.~Chapter 9 of \cite{ermentrout2010} as well as \cite{bertram2017} for relevant methods).
In particular, it is natural to represent such a CPG model as a collection of units, each with its own active and quiescient states, and to focus on solutions in which at most one unit is in its active state at any given time.    On the other hand, it is not necessarily the case that all components of a CPG circuit have the capacity to oscillate individually in some input range.  One alternative to the coupled relaxation oscillator CPG model is provided by models featuring heteroclinic cycles, sometimes referred to as winnerless competition models \citep{may1975nonlinear,afraimovich2004heteroclinic}.
In such models, an extended passage near a saddle point yields the prolonged duration associated with each phase of a CPG output.  Here, we no longer have a fast-slow timescale splitting and decomposition of the phase space into projections for individual units.
A third modeling framework for multi-phase neuronal rhythms, which also lacks a timescale splitting, is offered by the class of competitive, threshold-linear networks \citep{morrison2016diversity}.  Inhibitory interactions between threshold-linear units can yield oscillations without stable fixed points, as is suitable for functional CPG dynamics, and this modeling framework has been proposed as being convenient for mathematical tractability.

In each of these modeling frameworks, we ask how the architecture influences the controllability of the pattern of activation.
Specifically, we analyze the sensitivity of the individual activation phases comprising the overall pattern of activity in response to the tuning of an excitatory, tonic driving current external to -- that is, not generated by or affected by -- the activity of the circuit.  In all cases, we use a recently developed mathematical tool, the {\em local timing response curve} (lTRC) \citep{wang2021shape}.
The lTRC provides a first-order (linear) approximation for the effect of a parametric perturbation on the time it takes for a trajectory to pass through a given region.
It decomposes the effect into a component affecting the trajectory's entry position into the region, an impact on the vector field within the region, and an influence on its exit from the region.
Importantly for CPG models, this idea applies to vector fields that lack smoothness, or even feature discontinuities, at switching surfaces within the phase space.
The lTRC analysis provides a mathematically grounded numerical quantification  of the effects of a specific parameter variation on rhythm period and phase durations, which we can represent in terms of entrance, within-region, and exit effects.
In several cases, we supplement the lTRC calculation with mathematical analysis based on timescale decomposition or other ideas, to obtain additional insights about the dynamical mechanisms underlying the observed effects.
Our findings yield important information for future work modeling neuronal generation of multi-phase rhythms, in that they provide insights about the control properties of each framework that we consider, which can guide the selection of a modeling framework to match experimentally observed circuit properties.

In all cases, for concreteness, we consider three-phase, or triphasic, rhythms.
This choice is motivated by the common appearance of triphasic rhythms in CPG outputs, including those for mammalian respiration, tripod insect gaits, escape swimming in mollusks, scratching in turtles, feeding in sea slugs, and digestion in crustaceans \citep{marder2007,smith2007,buschges2008,daun2011,hao2011,wojcik2014,shaw2015significance}.
We organize the rest of our paper as follows.
A brief description of the modeling frameworks we consider and the related phase-transition mechanisms is given in section~\ref{sec:model_systems}.
Section~\ref{sec:Phase-duration_sensitivity} presents the main analysis and results on the controllability of phase durations in each phase transition mechanism.
Last, we summarize the mapping that we obtain between model classes and control properties in section \ref{sec:Discussion}, where we also discuss limitations, connections to previous literature, and possible implications of our results for biology and for future work.

\section{Model systems}
\label{sec:model_systems}

We will consider three classes of three-unit model systems that generate periodic oscillations in which units take turns activating, or exhibiting an epoch of elevated voltage, in a fixed sequence, with no two active at the same time.  
For reference, we label the units as 1, 2, 3, although all three will be identical. 
We refer to phase $i$ as the segment within one period of such a solution during which unit $i$ is active, and we use the notation $t_i$ to denote the duration of phase $i$.
The notation $\Delta t_i$ represents the change of $t_i$ in response to a perturbation of the control parameters.

\subsection{Relaxation oscillator circuit}

A vast array of planar dynamical systems, including many neuronal models, can produce relaxation oscillations.
In the neuronal case, the key ingredients are typically a fast inward current and a slow negative feedback, which can be the inactivation of the inward current or the activation of an outward current.
From a mathematical perspective, all such models that produce a transition from a stable hyperpolarized state to oscillations via an initial Andronov-Hopf (AH) bifurcation and a subsequent transition to a stable depolarized state via another   AH bifurcation, as a single input current parameter is varied, are equivalent (up to criticality of the AH bifurcations)
\citep{izhikevich2007dynamical}.
When such units are coupled together synaptically and produce sequential oscillations, however, their responses to parameter tuning can depend on the nature of the mathematical mechanism by which the neuronal activations alternate \citep{daun2009control,rubin2009multiple}.  
Thus, in this work, we consider a specific choice of neuronal relaxation oscillator model and we vary parameters within the context of that model to realize and compare several different transition mechanisms.

\paragraph{Equations}

For our individual neuron model, we follow past works on CPGs that used a slowly inactivating persistent sodium current ($I_\text{NaP}$), as has been characterized in respiratory CPG neurons \citep{butera1999models1,butera1999models2}, to support oscillations (e.g., \cite{rybak2006modelling,daun2009control,rubin2012explicit}).
The full circuit model includes mutual inhibitory coupling among three units.  We make the assumption that the coupling strength responds instantaneously to changes in voltage, which simplifies the analysis and represents a reasonable approximation because the synaptic conductance rise and decay rates are much faster than the timescale of $I_\text{NaP}$ (de)inactivation and which has been successfully used in previous analysis (see \cite{daun2009control,rubin2012explicit}).

The specific model equations that we consider are modified from a three-component respiratory CPG model \citep{rubin2012explicit}:
\begin{equation}
\label{eq:relaxation_equations}
\begin{split}
    v_1'=&(F(v_1,h_1)-g_\text{I}(b_{21}S_\infty(v_2)+b_{31}S_\infty(v_3))(v_1-V_\text{I})-g_\text{E}d_1(v_1-V_\text{E}))/C,\\
    v_2'=&(F(v_2,h_2)-g_\text{I}(b_{12}S_\infty(v_1)+b_{32}S_\infty(v_3))(v_2-V_\text{I})-g_\text{E}d_2(v_2-V_\text{E}))/C,\\
    v_3'=&(F(v_3,h_3)-g_\text{I}(b_{13}S_\infty(v_1)+b_{23}S_\infty(v_2))(v_3-V_\text{I})-g_\text{E}d_3(v_3-V_\text{E}))/C,\\
    h_1'=&(h_\infty(v_1)-h_1)/\tau_h(v_1),\\
    h_2'=&(h_\infty(v_2)-h_2)/\tau_h(v_2),\\
    h_3'=&(h_\infty(v_3)-h_3)/\tau_h(v_3).
\end{split}
\end{equation}
In system (\ref{eq:relaxation_equations}), $F$ describes a persistent sodium current and a leak current,
$$F(v,h)=-g_\text{NaP}m_{p\infty}(v)h(v-V_\text{Na})-g_\text{L}(v-V_\text{L}).$$
The coupling function $S_\infty$ is given by a monotone increasing function taking values in $[0,1]$,
$$S_\infty(v)=1/(1+\exp{((v-\theta_\text{I})/\sigma_\text{I})}),$$
where $\theta_\text{I}$ represents the synaptic threshold. This function closely approximates a Heaviside step function when $\sigma_\text{I}$ is small, and it is multiplied by a coupling strength $b_{ij}$ for $i,j \in \{1,2,3 \}$ each time it appears.
The final term, $g_\text{E}d_i(v_i-V_\text{E})$, in each voltage equation represents a tonic
excitatory drive, and we consider the strength factors $d_i$ as control parameters.
In the equations for gating variables,
$\tau_h$ is the time constant for the sodium conductance inactivation, given by
$$\tau_h^{-1}(v)=\epsilon\cosh{((v-\theta_h)/(2\sigma_h))},$$
where $0<\epsilon\ll1.$
Additional details about the functions in system \eqref{eq:relaxation_equations}, as well as parameter values used for simulations, are given in Appendix~\ref{App:model_details}.

Figure~\ref{fig:IR_solution} shows a typical solution of system~\eqref{eq:relaxation_equations}, where each cell undergoes relaxation-oscillator-like dynamics.
Because the coupling between units in (\ref{eq:relaxation_equations}) only occurs through the functions $S_{\infty}(v_i)$, which transition rapidly between values near 0 and 1, we can represent the solution as a collection of three phase plane projections, one to each coordinate pair $(v_i,h_i)$.  In each phase plane, since $\epsilon$ is small, the projected trajectory will evolve along a branch of the $v_i$-nullcline, except for rapid jumps between branches at the fold points where pairs of branches meet and end. 
A rapid change in an $S_{\infty}$ term in a neuron's voltage equation yields a rapid change in its $v$-nullcline position and corresponding jump to a new branch.
An example of the $(v_1,h_1)$ projection is shown in Figure~\ref{fig:IR_solution}(b).
There, we display two distinct $v_1$-nullclines. Recall that the $v_1$-nullsurface in the 6-dimensional phase space of system (\ref{eq:relaxation_equations}) defined by $v_1'=0$. 
This becomes a one-dimensional curve when projected to $(v_1,h_1)$-space for any fixed $(v_2,v_3)$, and the position of this curve depends on the values assigned to $v_2$ and $v_3$. 
Because the function $S_{\infty}(v_i)$ is steep and we consider solutions in which at most one cell is active at a time, most of the time we have $(S_{\infty}(v_2),S_{\infty}(v_3))$ within a small neighborhood of one of the three points $\{ (0,0), (0,1), (1,0) \}$. Hence, three positions of the projected $v_1$-nullsurface are most relevant for illustrating trajectory behavior on the $(v_1,h_1)$-plane, and these reduce to two because we take $b_{21}=b_{31}$. 
We refer to the projected $v_1$-nullsurface for $(0,0)$ as the active or uninhibited $v_1$-nullcline and to that for $(1,0)$, and equivalently for $(0,1)$, as the inhibited $v_1$-nullcline.

In the solutions we consider, there is always exactly one cell active at each time (except during jumps), which inhibits the other two cells.
As parameters are varied, a number of different activation patterns can be generated, e.g. (132), (1232), (13123), etc. (see more patterns in \cite{rubin2012explicit}).
In the rest of the paper, we consider the symmetric (123) firing pattern of the system, also known as \emph{splay states}, in which the three cells alternate activating in the activation order $1\rightarrow2\rightarrow3\rightarrow1\rightarrow2\rightarrow3\rightarrow\cdots$.
Because all system parameters are the same for all neurons in the simulation shown in Figure~\ref{fig:IR_solution}, a splay state occurs.  Thus, the phase plane projections for the other two units are identical to those shown for $(v_1,h_1)$; although the solution time courses are phase-shifted (Figure~\ref{fig:IR_solution} (a)), that difference does not show up in the phase plane (Figure~\ref{fig:IR_solution}(b)), since time is not explicitly represented there.

\begin{figure}
\centering
\includegraphics[width=15.5cm]{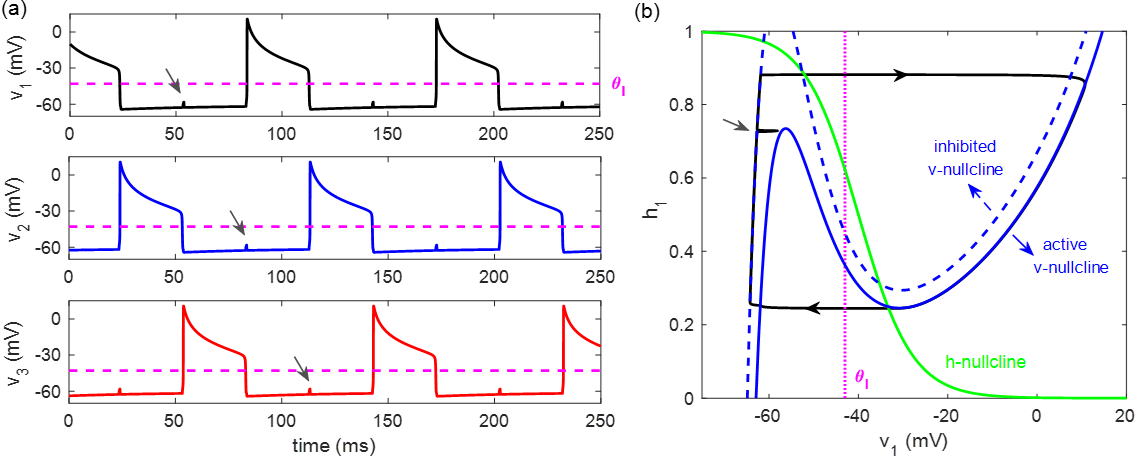}%
\caption{\label{fig:IR_solution} A typical solution of system \eqref{eq:relaxation_equations}, which features an intrinsic release mechanism. \textbf{(a)}: Voltage versus time for the three cells. The three cells evolve in a symmetric way and take turns activating. Magenta dashed horizontal line: synaptic activation threshold $\theta_\text{I}$. \textbf{(b)}: $(v_1, h_1)$ phase plane of the solution. Black solid curve: cell 1 limit cycle trajectory; green solid curve: $h_1$-nullcline; blue dashed curve: inhibited $v_1$-nullcline; blue solid curve: active $v_1$-nullcline; magenta dashed vertical line: synaptic activation threshold $\theta_\text{I}$. Phase planes for $(v_2,h_2)$ and $(v_3,h_3)$ are the same as $(v_1,h_1)$.
Note the transient upward voltage deflections (grey arrows) during the brief interval in which the silent cell is released from inhibition but then suppressed due to the faster activation of the other silent cell (cf. \cite{rubin2012explicit,park2022}).}
\end{figure}

\paragraph{Phase transition mechanisms}

\cite{skinner1994mechanisms} 
recognized that in half-center oscillators, defined as pairs of neurons that generate alternating oscillations when coupled by reciprocal inhibition, switches of which cell is active can occur through four different transition mechanisms: \emph{intrinsic release}, \emph{synaptic release}, \emph{intrinsic escape} and \emph{synaptic escape} (see also \cite{wang1992alternating,zhang2013phase,yu2021dynamical}).
These concepts extend naturally to larger networks with reciprocally inhibitory coupling \citep{rubin2009multiple,sakurai2022bursting}.
The intrinsic properties of the neurons and the properties of the synapses between them, especially the position of the synaptic threshold, determine the fundamental differences of the four mechanisms.

\paragraph{Intrinsic release} Figure~\ref{fig:IR_solution} illustrates the dynamics of the intrinsic release mechanism.
During the network oscillation, the active cell reaches the lower right knee of its active (uninhibited) $v$-nullcline and jumps down.
As it does so, its voltage crosses through the synaptic threshold $\theta_\text{I}$,  releasing the suppressed cell with the largest $h$-value from inhibition.
Although the other suppressed cell's voltage also transiently increases, it is quickly suppressed by the activation of the neuron with the largest $h$ (see grey arrows in Figure~\ref{fig:IR_solution}). The formerly active cell becomes inhibited by the newly-active cell and thus completes its jump down on the left branch of its inhibited $v$-nullcline.

\begin{figure}
\centering
\includegraphics[width=15.5cm]{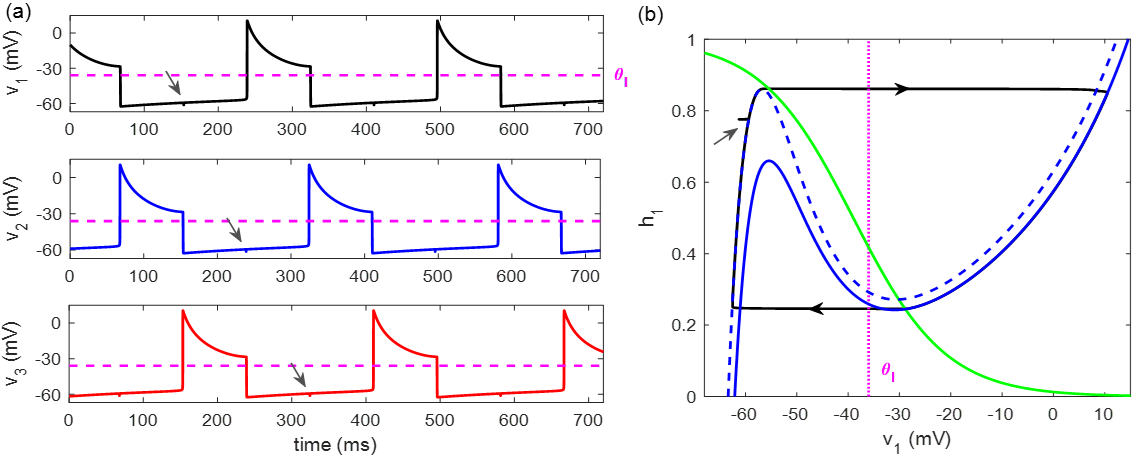}
\caption{\label{fig:IE_solution_symmetric} Dynamics of the intrinsic escape mechanism. Colors as in Figure~\ref{fig:IR_solution}. When the inhibited cell reaches the upper left knee of its $v$-nullcline, it jumps up to the active $v$-nullcline and inhibits the other two cells.
Note the transient downward voltage deflection (grey arrows) during the brief interval in which the cell is inhibited by both of the other cells in the network, which we call a ``double inhibition notch".}
\end{figure}

\paragraph{Intrinsic escape}
In the intrinsic escape mechanism, the oscillations are controlled by the suppressed cell escaping from inhibition when it reaches the upper left knee of its $v$-nullcline. Figure~\ref{fig:IE_solution_symmetric} shows the time course of a splay state with all transitions by intrinsic escape and the phase plane projection of the trajectory to $(v_1,h_1)$-space.  At first glance, the projection may look identical to the intrinsic release case.  Note, however, that cell 1 reaches the left knee of its inhibited $v$-nullcline (dashed blue) and escapes, or jumps up, from there despite still being inhibited.   Moreover, once cell 1 is active, it converges to a stable critical point where its active $v$-nullcline (solid blue) intersects its $h$-nullcline (green) at $v>\theta_\text{I}$.  The subsequent jump-down of cell 1 happens only because cell 2 manages to reach its left knee and escape.
Different from the intrinsic release in which a transient increase occurs for the voltage of the suppressed cell, in the intrinsic escape case a transient decrease exists due to double inhibition from the other two active cells (see grey arrows in Figure~\ref{fig:IE_solution_symmetric}), which we call a ``double inhibition notch".

\begin{figure}
\centering
\includegraphics[width=15.5cm]{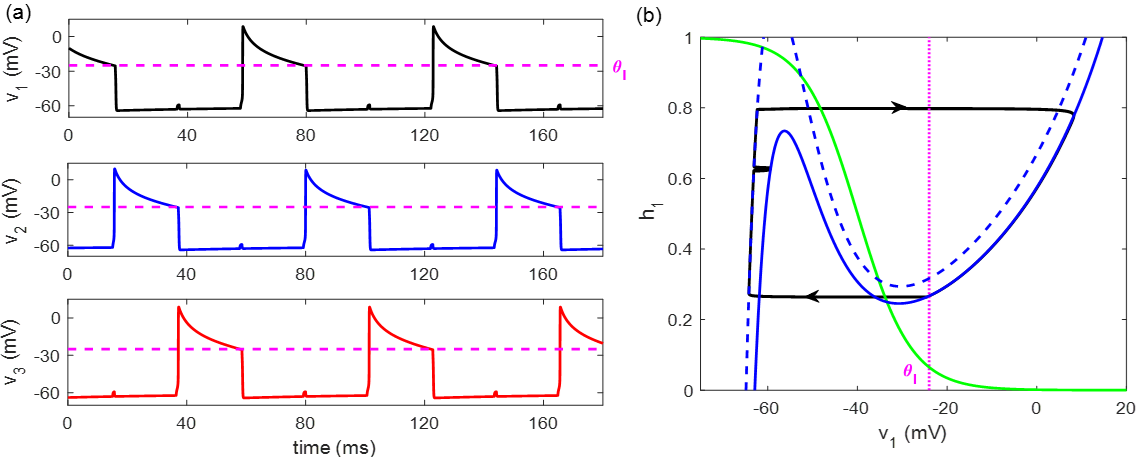}%
\caption{\label{fig:SR_solution} Dynamics of the synaptic release mechanism. Colors as in Figure~\ref{fig:IR_solution}. The synaptic threshold $\theta_\text{I}$ intersects the right branch of the active voltage nullcline. The release occurs when the active cell's voltage declines to the synaptic threshold.}
\end{figure}

\begin{figure}
\centering
\includegraphics[width=15.5cm]{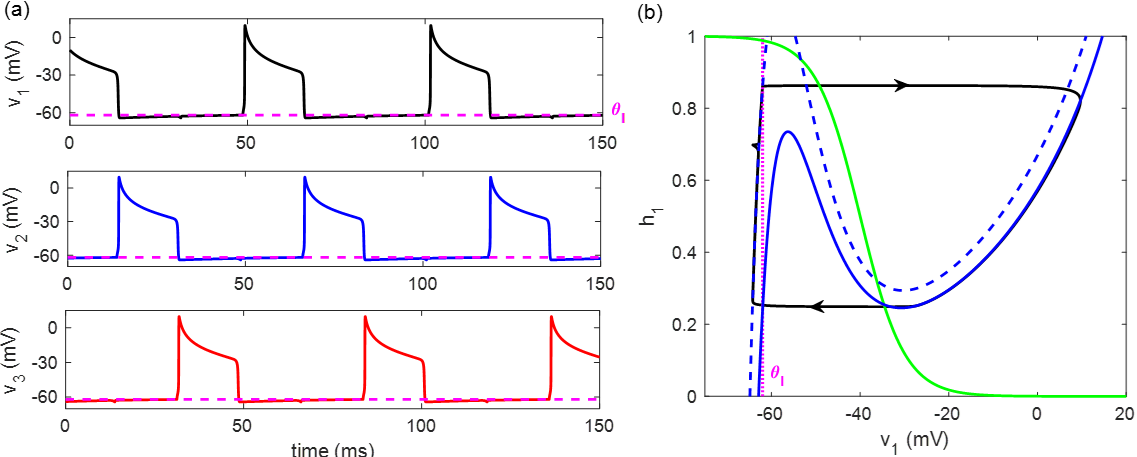}
\caption{\label{fig:SE_solution} Dynamics of the synaptic escape mechanism. Colors as in Figure~\ref{fig:IR_solution}. The synaptic threshold $\theta_\text{I}$ intersects the left branch of the inhibited voltage nullcline. When the suppressed cell passes the synaptic threshold, it initiates a phase transition.}
\end{figure}

\paragraph{Synaptic release}
The synaptic mechanisms come into operation when the synaptic threshold $\theta_\text{I}$ intersects one of the outer branches of the voltage nullclines, rather than the middle branch.  In this case,
the cells' transitions between the active and
suppressed states occur due to synaptic dynamics.
Figure~\ref{fig:SR_solution} illustrates an example of the synaptic release mechanism, where $\theta_\text{I}$ lies on the right branch of the active $v$-nullcline.
The jumps occur when the active cell reaches the synaptic threshold voltage.  When this happens, it no longer fully inhibits the suppressed cells, and the one with largest $h$ can activate. When that cell's voltage jumps up, it crosses the synaptic threshold and inhibits the active cell, which in turn jumps down, starting from $v \approx \theta_\text{I}$ since all of this action happens on the fast timescale.
Note in Figure~\ref{fig:SR_solution}(b) that the resulting trajectory projection does not pass through either the local maximum or minimum of the $v$-nullcline.
\paragraph{Synaptic escape}
In this mechanism, the synaptic threshold intersects the left branch of the inhibited nullcline, as shown in Figure~\ref{fig:SE_solution}. The jumps are determined by when an inhibited cell's voltage crosses through the synaptic threshold.  When this occurs, the active cell becomes inhibited and jumps down, which turns off the inhibition to the suppressed cells and allows the inhibited cell with the newly supra-threshold voltage to activate.

\subsection{Heteroclinic cycling model}
\label{ssec:heteroclinic_cycling_model}

\begin{figure}
\centering
\includegraphics[width=15.5cm]{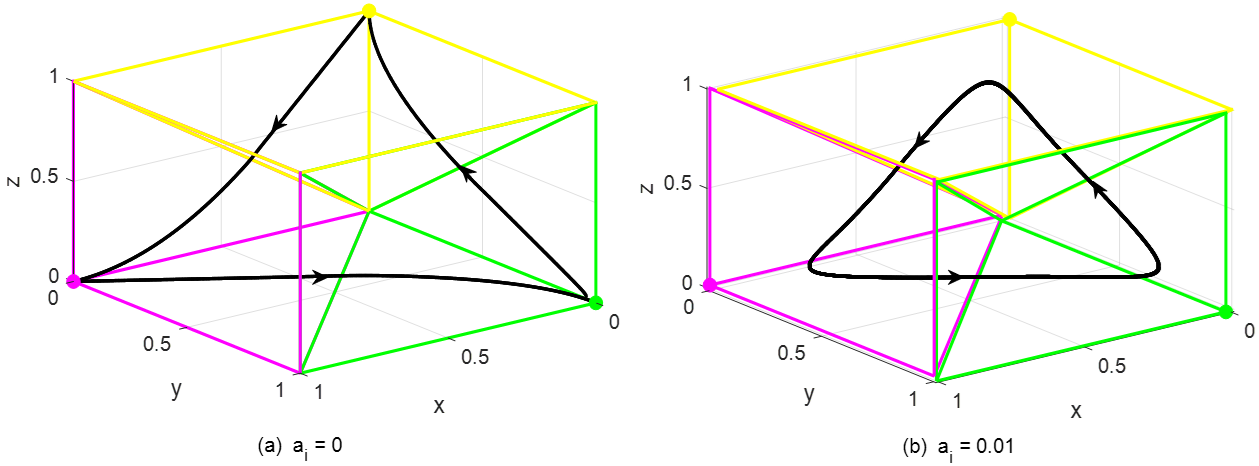}
\caption{\label{fig:Aplysia_solutions} Piecewise linear model of \textit{Aplysia} feeding system (\ref{eq:Aplysia_equations}) for $\rho=3$ and two values of endogenous neural excitation parameters $a_i$.
The domain is divided into three equal pyramidal regions, with boundaries colored magenta, green and yellow, respectively.
A saddle fixed point (solid dot) lies on the corner of each region.
The magenta, green and yellow regions are the respective active regions for pool $x$, $y$ and $z$.
Arrows mark the direction of flow.
\textbf{(a)} When $a_1=a_2=a_3=0$, the system contains a stable heteroclinic cycle.  \textbf{(b)}
When $a_1=a_2=a_3=0.01$, the system contains a stable limit cycle.
Note that in this case, the critical points are no longer exactly at the corners of the unit cube.}
\end{figure}

\cite{shaw2015significance} and \cite{lyttle2017robustness} introduced a firing-rate model for the feeding pattern generator of the marine mollusk \textit{Aplysia californica}, adapted from May and Leonard's three-population heteroclinic cycling model \citep{may1975nonlinear}.
Here we study a three-dimensional piecewise linear system adapted from \cite{park2018infinitesimal} that was introduced as a piecewise linear simplification of the (smooth) heteroclinic cycling model.
This system takes the form
\begin{equation}
\label{eq:Aplysia_equations}
\begin{split}
    \frac{d}{dt}\left(\begin{matrix}x\\y\\z\end{matrix}\right)=\left\{\begin{aligned}
    &\left.\begin{aligned}
    &\;1-x-(y+a_1)\rho\\
    &\;y+a_2\\
    &\;(z-a_3)(1-\rho)\end{aligned}\right.\qquad \mbox{if} \; \; x\geq y+\frac{a_1+a_2}{2},\;x\geq z-\frac{a_1+a_3}{2}\\
    &\\
    &\left.\begin{aligned}&\;(x-a_1)(1-\rho)\\
    &\;1-y-(z+a_2)\rho\\
    &\;z+a_3\end{aligned}\right.\qquad \mbox{if} \; \; y>x-\frac{a_1+a_2}{2},\;y\geq z+\frac{a_2+a_3}{2}\\
    &\\
    &\left.\begin{aligned}&\;x+a_1\\
    &\;(y-a_2)(1-\rho)\\
    &\;1-z-(x+a_3)\rho\end{aligned}\right.\qquad \mbox{if} \; \;
    z>x+\frac{a_1+a_3}{2},\;z>y-\frac{a_2+a_3}{2}.
    \end{aligned}\right.
\end{split}
\end{equation}
This system incorporates firing rates of three neural pools --- the ``protraction-open" pool $x$, the ``protraction-closed" pool $y$, and the ``retraction" pool $z$.
Parameter $a_i$ represents intrinsic neural excitation to each pool, and $\rho$ is a coupling constant representing inhibition between the neural pools.
The domain of system \eqref{eq:Aplysia_equations} is divided into three equal rectangular pyramids that together comprise the unit cube, as shown in Figure~\ref{fig:Aplysia_solutions}.
When all $a_i=0$, the intrinsic neural dynamics contain a stable heteroclinic cycle connecting saddles at $(1, 0, 0)$, $(0, 1, 0)$, and $(0, 0, 1)$, see Figure~\ref{fig:Aplysia_solutions}(a). When some $a_i$ are small and positive, the heteroclinic cycle is broken and a stable limit cycle arises, see Figure~\ref{fig:Aplysia_solutions}(b).

\subsection{Competitive threshold-linear network}

The third framework we consider is a competitive threshold-linear three-unit network governed by the following ODE system \citep{morrison2016diversity}:
\begin{equation}
\label{eq:threshold_linear_equations}
    \dot{x}_i=-x_i+\left[\sum_{j=1}^3W_{ij}x_j + \theta_i \right]_+,\quad i\in\{1,2,3\}.
 \end{equation}
The connectivity matrix $W=[W_{ij}]$ in (\ref{eq:threshold_linear_equations}) gives directed connection strengths between pairs of nodes, the parameter $\theta_i$ is an external drive to node $i$, and the 
nonlinearity $[\cdot]_+$ is given by $[y]_+ = \max\{y,0\}$.

Assume the connectivity matrix $W$ to be
\begin{equation*}
    W_{ij}=\left\{\begin{aligned}
    &0,&\text{if}\;j=i\\
    &-1+\epsilon,&\text{if}\;j\rightarrow i\\
    &-1-\delta,&\text{if}\;j\nrightarrow i
    \end{aligned}\right.
\end{equation*}
where $\delta>0$ and $0<\epsilon<1$.
It is shown in \cite{morrison2016diversity} that in a unidirectional coupled network, if $$\epsilon<\frac{\delta}{1+\delta},$$
then the network has bounded activity and no stable fixed points.
Specifically, to have the nodes oscillate in the cyclic order $(123)$, we can take
\begin{equation*}
    W=\left(\begin{matrix}0&-1-\delta&-1+\epsilon\\-1+\epsilon&0&-1-\delta\\-1-\delta&-1+\epsilon&0\end{matrix}\right),
\end{equation*}
where $\epsilon=0.25$ and $\delta=0.5$ to satisfy the given inequality.

If all of the drive terms $\theta_i$ are set to 0, then the system trajectories converge to the stable fixed point $(0,0,0)$.
When the $\theta_i$ are all positive, the solution trajectory always converges to a stable limit cycle, which surrounds an unstable fixed point.
This oscillation mechanism differs from that in the heteroclinic cycling model (\S\ref{ssec:heteroclinic_cycling_model}), where the periodic orbit spends its time in small neighborhoods of invariant manifolds associated with equilibria.

\begin{figure}
\centering
\includegraphics[width=15.5cm]{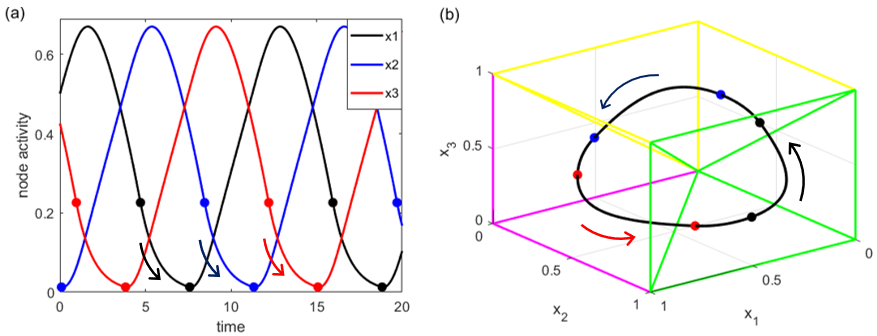}%
\caption{\label{fig:TL_reference} Activity dynamics of the three nodes in the competitive threshold-linear network \eqref{eq:threshold_linear_equations} with $\theta_1=\theta_2=\theta_3=1$. \textbf{(a)}: Time courses of nodes $x_1$ (black), $x_2$ (blue) and $x_3$ (red). \textbf{(b)}: Trajectory in the 3D cubic domain. Node $x_1$ is active in the magenta region; $x_2$ is active in the green region; $x_3$ is active in the yellow region.
Arrows mark the direction of flow.
Because of the rectifying nonlinearities, the trajectory passes through a series of six switching surfaces per periodic orbit.  Black, blue, and red solid dots in panels (a) and (b) mark nonsmooth points of the dynamics.}
\end{figure}

To make a clear comparison with the heteroclinic cycling model \eqref{eq:Aplysia_equations}, we again partition the unit cube domain into three equal pyramids and define the active region for each node to be
\begin{align*}
    \left\{\begin{aligned}
    &x_1\;\text{active},&\text{if}\;x_1\geq x_2,\,x_1\geq x_3\\
    &x_2\;\text{active},&\text{if}\;x_2> x_1,\,x_2\geq x_3\\
    &x_3\;\text{active},&\text{if}\;x_3> x_1,\,x_3> x_2.
    \end{aligned}\right.
\end{align*}
Figure~\ref{fig:TL_reference} is a reference figure showing the periodic trajectory in the default symmetric case $\theta_1=\theta_2=\theta_3=1$.
The three nodes activate in the cyclic order $(123)$ with equal durations spent in each region.

\section{Phase-duration sensitivity}
\label{sec:Phase-duration_sensitivity}

A functional CPG should produce a periodic oscillation that is stable to perturbations and can be tuned in response to environmental conditions and metabolic demands.  A natural expectation might be that each component of a CPG circuit would receive an independent control signal, which could be adjusted to modulate its active phase duration and amplitude.  This view, however, overlooks the reality that when a dynamical system produces a periodic behavior, the coupling among the system components may interfere with or prevent such distributed control (e.g., \cite{daun2009control}).  Thus, although there may be an independent input to each network component, as is the case with the models presented in the previous section, tuning of these inputs may affect the properties of multiple phases of the output rhythm.

Therefore, in this section, we study the responses of periodic solutions in each model class to variations in a control parameter.  Since we consider the case of symmetric networks, in that the dynamics of all network components and the values of all coupling strengths are identical, we (without loss of generality) vary the control parameter to component one in each model and study the effect of this variation on all three phase durations within the perturbed oscillation.  We then explain the observed responses via two approaches: a quantitative analysis using the \textit{local timing response curve} (lTRC) and various model-specific qualitative analyses based on other mathematical tools.

The lTRC calculation developed in \cite{wang2021shape} quantifies the
timing sensitivity of a limit cycle within any given \emph{local} region bounded between specific Poincar\'{e} sections in phase space, by decomposing the response to a static perturbation into components that relate to entry section crossing, exit section crossing, and alteration of the vector field within the region.
Consider an ordinary differential equation model
$$\frac{dx}{dt} = F(x(t),\mu)$$
for $x(t)$ in a phase space $\Omega \subset \mathbb{R}^n$, $n\ge 2$, and a parameter $\mu \in \mathbb{R}$.
Without loss of generality, we take $\mu=0$ as a baseline or unperturbed case and let $F(x(t))$ denote $F(x(t),0)$.
Suppose that in the baseline case the system supports a stable periodic solution, or limit cycle, $\gamma$.
Suppose that $\gamma$ passes through a region or subdomain $R \subset \Omega$, with entry point $x^\text{in}$ and exit point $x^\text{out}$.
For each $x \in \gamma \cap R$, let $\Gamma(x)$ be the time it takes for the trajectory that originates at $x$ to exit from $R$.
The lTRC is defined to be the gradient of $\Gamma(x)$, $\eta(t)=\nabla\Gamma(x(t))$,
which satisfies the adjoint equation
$$\frac{d\eta}{dt}=-DF(\gamma(t))^T\eta,$$
where $DF$ denotes the Jacobian of the vector field $F$.
At the exit point $x^\text{out}$ of region $R$,  $\eta$ satisfies the boundary condition
$$\eta(x^\text{out})=-\frac{n^\text{out}}{(n^{\text{out}})^T%
F(x^\text{out})},$$
where $n^{\text{out}}$ is a normal vector of the exit boundary surface at $x^\text{out}$ (see more details about the lTRC in Appendix~\ref{appssec:Details of lTRCs}).

Using the lTRC, \cite{wang2021shape} provide a formula for analytically calculating the \emph{first-order} or \emph{linear} approximation for the change in the duration a trajectory spends in region $R$, induced by a small change in $\mu$.
That is, writing $\Gamma_{\mu}(x)$ for the time to exit $R$ from point $x$ under the perturbation $\mu$, we have $\Gamma_\mu(x_\mu^\text{in})=\Gamma(x^\text{in})+\mu T_1+O(\mu^2)$, with:
\begin{equation}
\label{eq:T1_fixed_boundary}
  T_1=\eta(x^\text{in})\cdot\frac{\partial x_{\mu}^\text{in}}{\partial\mu}\bigg|_{\mu=0}+\int_{t^\text{in}}^{t^\text{out}}\eta(\gamma(t))\cdot\frac{\partial F_{\mu}(\gamma(t))}{\partial\mu}\bigg|_{\mu=0}\,dt,
\end{equation}
where $x^\text{in},x_{\mu}^\text{in}$ represent the unperturbed entry point and the perturbed entry point to region $R$, respectively, and $F_{\mu}(x(t)) := F(x(t),\mu)$.
The first-order duration shift $T_1$ given by \eqref{eq:T1_fixed_boundary} consists of two terms: the first term arises from the impact of the perturbation on the entry point to the region; the integral term shows the impact of the perturbation on the vector field.

Note that in \cite{wang2021shape,Wang2022variational}, boundary surfaces of region $R$ were assumed to be fixed in the derivation of (\ref{eq:T1_fixed_boundary}).
When considering boundary surfaces that may themselves change position  as control parameters vary (e.g., the boundaries of each region in the heteroclinic cycling model \eqref{eq:Aplysia_equations} shift as parameters $a_i$ are varied), a third term for $T_1$ arises, accounting for the impact of the perturbation on the exit point:
\begin{equation}
\begin{split}
\label{eq:T1_shifting_boundary}
    T_1&=\eta(x^\text{in})\cdot\frac{\partial x_{\mu}^\text{in}}{\partial\mu}\bigg|_{\mu=0}-\eta(x^\text{out})\cdot\frac{\partial x_{\mu}^\text{out}}{\partial\mu}\bigg|_{\mu=0}+\int_{t^\text{in}}^{t^\text{out}}\eta(\gamma(t))\cdot\frac{\partial F_{\mu}(\gamma(t))}{\partial\mu}\bigg|_{\mu=0}\,dt\\
    &=\int_{t^\text{in}}^{t^\text{out}}\underbrace{\delta(t-t^\text{in})\cdot\eta(x^\text{in})\cdot\frac{\partial x_{\mu}^\text{in}}{\partial\mu}\bigg|_{\mu=0}}_\text{term A}
    -\underbrace{\delta(t-t^\text{out})\cdot\eta(x^\text{out})\cdot\frac{\partial x_{\mu}^\text{out}}{\partial\mu}\bigg|_{\mu=0}}_\text{term B}
    +\underbrace{\eta(\gamma(t))\cdot\frac{\partial F_{\mu}(\gamma(t))}{\partial\mu}\bigg|_{\mu=0}}_\text{term C}\,dt,
\end{split}
\end{equation}
where $\delta(\cdot)$ is the Dirac delta function.
When the exit boundary is fixed,
$\eta(x^\text{out})=\nabla\Gamma(x^\text{out})$ is parallel to $n^\text{out},$ the vector normal to the exit boundary.
In this case, $n^\text{out} \perp\partial x_\mu^\text{out}/\partial\mu$; therefore, term B is identically zero and \eqref{eq:T1_shifting_boundary} naturally reduces to \eqref{eq:T1_fixed_boundary}.
The derivation of our generalized formula \eqref{eq:T1_shifting_boundary} for the shifting-boundary case is given in Appendix~\ref{appssec:Derivation of generalized T_1 formula}.

With the linear duration shift $T_1$ calculated from equation \eqref{eq:T1_fixed_boundary} or equation \eqref{eq:T1_shifting_boundary}, we can obtain the phase duration change to the first order, given by $\mu\times T_1$.
Numerically, in the calculation of term A and term B in equation \eqref{eq:T1_shifting_boundary} we approximate the linear shift in the entry and exit positions on Poincar\'e sections by
$$\frac{\partial x_\mu^\text{in}}{\partial\mu}\bigg|_{\mu=0}\approx\frac{x_\mu^\text{in}-x^\text{in}}{\mu},\qquad\frac{\partial x_\mu^\text{out}}{\partial\mu}\bigg|_{\mu=0}\approx\frac{x_\mu^\text{out}-x^\text{out}}{\mu}.$$
Note that $T_1>0$ (resp. $T_1<0$) if a positive perturbation increases (resp. decreases) the phase duration.
In the following subsections, we use the lTRC calculation to study the phase duration sensitivity to a specific perturbation (with a comparison to direct numerical simulation), and we supplement the lTRC analysis with other mathematical approaches such as fast-slow decomposition and vector field analysis.

\subsection{Relaxation oscillator circuit}

In the relaxation oscillator circuit \eqref{eq:relaxation_equations}, we start from the  symmetric case ($d_1=d_2=d_3=1$) and apply a small sustained perturbation $\mu$ to $d_1$ (i.e., $d_1\rightarrow d_1+\mu$).
We immediately run into a complication in this case:  past work has shown that the response of such models to tuning a constant input parameter depends on the phase transition mechanism operating in the unperturbed periodic orbit \citep{daun2009control,rubin2009multiple}.
We consider the four transition mechanisms discussed by \cite{skinner1994mechanisms} and reviewed in the previous section: intrinsic and synaptic release and escape.
We display and compare the sensitivities of the active phase durations for three of the four cases; to our surprise, we find that, unlike in a two-cell network, the symmetric periodic orbit with transitions by intrinsic escape is unstable to small perturbations in the three-cell network, so we conclude this section with commentary on this instability (with additional analysis of the instability in Appendix~\ref{app:IE_instability}).

\begin{figure}
\centering
\includegraphics[width=14cm]{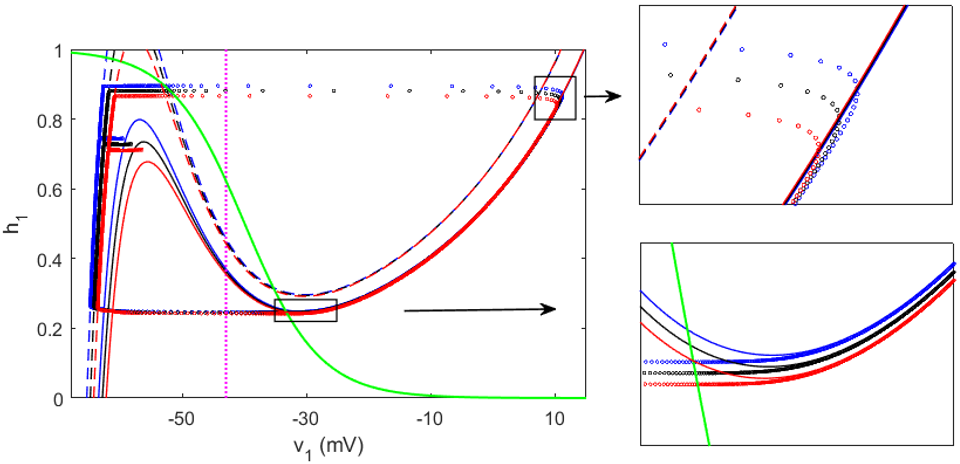}
\caption{\label{fig:IR_phase1} Projections to the phase plane for cell 1 in the intrinsic release case for $d_1=0.7$ (blue), $d_1=1$ (black) and $d_1=1.3$ (red). The color of each trajectory is in line with the color of the corresponding $v_1$ nullcline; the $h_1$-nullcline appears in green.
Solid nullclines are relevant when cells 2 and 3 are inactive, while dashed nullclines arise when cell 2 or cell 3 is active.
Successive dots are equally spaced in time.
As $d_1$ increases, the $v_1$ nullclines are lowered, and consequently, the jump-down position (right knee) of cell 1 is lowered (see bottom rectangle).
The trajectories' proximity to the $h_1$-nullcline translates a small spatial dispersion into a large temporal effect, such that the prolonging of the active phase at jump-down
is much more significant than the  reduction occurring in the entry to the active phase (upper rectangle), far from the $h_1$-nullcline.}
\end{figure}

\begin{figure}
\centering
\includegraphics[width=14cm]{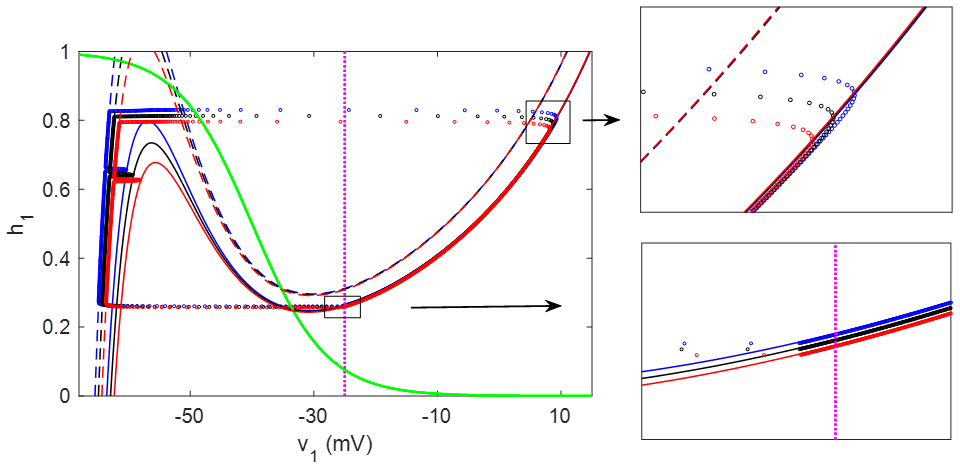}
\caption{\label{fig:SR_phase1} Projections to the phase plane for cell 1 in the synaptic release case for $d_1=0.7$ (blue), $d_1=1$ (black) and $d_1=1.3$ (red).
Colors are as in Figure~\ref{fig:IR_phase1}.
A positive  perturbation to $d_1$ results in a decrease in the $h_1$-coordinate of the jump-down point.
In contrast to the intrinsic release case (Figure~\ref{fig:IR_phase1}), the jump-down occurs with the $v_1$-coordinate invariant with respect to $d_1$ at  $v_1\approx\theta_\text{I}$, which is farther from the $h_1$-nullcline, such that the change in spatial position does not translate into as large  of an increase in the active phase duration.}
\end{figure}

The dominant effect of increasing the value of $d_1$ is that the active $v_1$-nullcline, along which cell 1 moves during the active phase, is lowered, as illustrated in Figures~\ref{fig:IR_phase1} and \ref{fig:SR_phase1}. For visual convenience, Figures~\ref{fig:IR_phase1}, \ref{fig:SR_phase1} show the trajectories with a perturbation of magnitude $|\mu|=0.3$ applied to $d_1=1$, but in our actual analysis and lTRC calculation we impose a smaller perturbation of magnitude $|\mu|=0.05$.

We define neuron $i$ as being in the active phase when $v_i > \theta_\text{I},$ the synaptic threshold. The sections $\{ v_i = \theta_\text{I} \}$ partition phase space into 8 regions, based on the various possible combinations of active and inactive neurons.
For each neuron, we are interested in its active phase duration, so for each $i \in \{ 1, 2, 3 \}$ we define the Poincar\'e sections
\begin{equation}
\label{eq:relaxation_Poincare_sections}
\begin{split}
    P_{i,\text{up}}&=\{v_i=\theta_\text{I},\;dv_i/dt>0\},\\
    P_{i,\text{down}}&=\{v_i=\theta_\text{I},\;dv_i/dt<0\}
\end{split}
\end{equation}
and we quantify the time from the passage of the trajectory through $P_{i,\text{up}}$ until its passage through $P_{i,\text{down}}$ for each $i$.
By equation \eqref{eq:T1_shifting_boundary}, for each region we can analytically calculate the linear shift in the duration of each active phase when a small static perturbation $\mu$ is applied to $d_1$.

\paragraph{Intrinsic release}
Table~\ref{tab:IR_duration_difference} presents the active phase duration changes for a solution featuring the intrinsic release transition  mechanism, obtained both by direct numerical simulation and by the lTRC calculation.
The analytical results show good agreement with the numerical simulation, with both revealing that an increase in $d_1$ prolongs the active phase of cell 1 and leaves those of cells 2 and 3 almost unchanged.

\begin{table}
\centering
\caption{\label{tab:IR_duration_difference} Intrinsic release: duration changes of the three individual active phases, computed by direct numerical simulation (``Simulated difference'') or lTRC calculation (``lTRC difference'').
Specifically, $\Delta t_i$ in the ``Simulated difference" column is given by the difference between the perturbed duration and unperturbed duration, and in the ``lTRC difference" it is a first-order approximation, given by $\mu T_1$ where $T_1$ is obtained from (4) or (5).
The durations in the default case ($d_1=d_2=d_3=1$) are $t_1=t_2=t_3=29.3227$. Phase 1 is prolonged in response to a positive $d_1$-perturbation, while phases 2 and 3 are insensitive. }
\begin{tabular}{|lcccclccc|}
\hline
Simulated difference&$\Delta t_1$&$\Delta t_2$&$\Delta t_3$&&lTRC difference&$\Delta t_1$&$\Delta t_2$&$\Delta t_3$\\
\hline
$\mu=0.05$&0.1118&0.0008&0.0009&&$\mu=0.05$&0.1033&0.0008&0.0008\\
$\mu=-0.05$&-0.1107&-0.0009&-0.0008&&$\mu=-0.05$&-0.1033&-0.0008&-0.0008\\
\hline
\end{tabular}
\end{table}

\begin{figure}
\centering
\includegraphics[width=15.5cm]{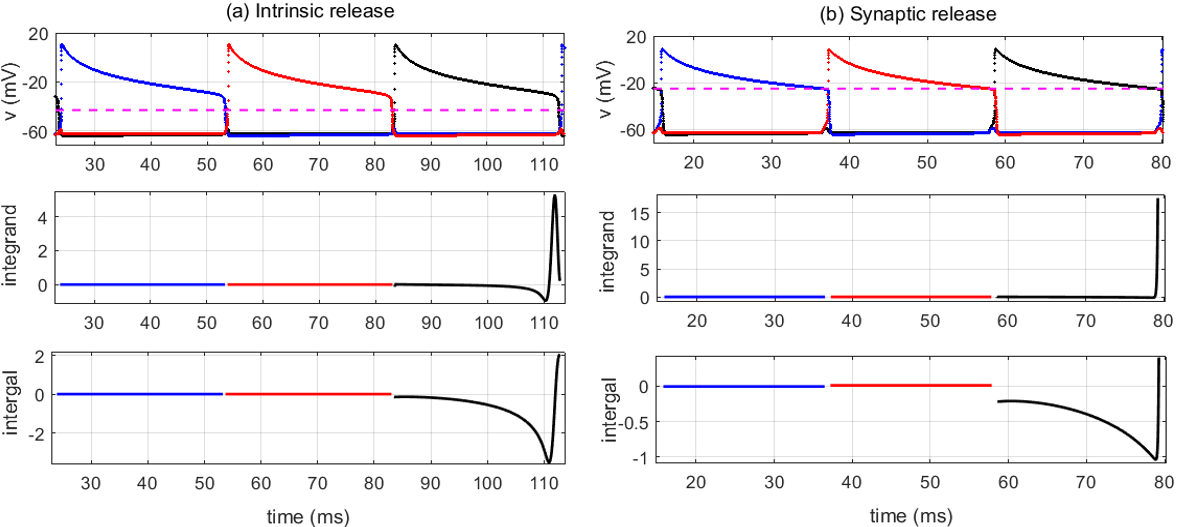}
\caption{\label{fig:lTRC_IR_SR} Timing responses for \textbf{(a)} intrinsic-release mechanism and \textbf{(b)} synaptic-release mechanism, measured by equation \eqref{eq:T1_shifting_boundary}.
Transitions into and out of each active phase are defined by \eqref{eq:relaxation_Poincare_sections}, and the respective timing characteristics in each phase are calculated.
Top panels: time course of membrane potentials over one period. Middle panels: integrand in $T_1$ at each time, given by \eqref{eq:integrand}.
Bottom panels: accumulated integral of $T_1$ (the area between the integrand curve and $x$-axis).
Black, blue and red traces correspond to when cell 1, cell 2 and cell 3, respectively, is active.
\textbf{Intrinsic release}: When a positive perturbation is applied to $d_1$, cell 1 (black) starts with a negative response upon entering the active phase at $t=83.49$; the negative response continues accumulating, until $t= 110.88$ (where the integrand is zero); near the end of the active phase, the integrand curve rapidly increases to large positive values, which cancels out the negative response accumulated before and finally ends up with a positive response.
Cells 2 and 3 react slightly to the perturbation only at their transition to the active state (at $t=53.71$ and $t=23.93$, respectively).
\textbf{Synaptic release}: Although the integrand curve at the end of the phase 1 transiently increases to a value larger than for intrinsic release, its accumulated effect (bottom panels) is much smaller than for intrinsic release, leading to less variation in the active duration of cell 1.}
\end{figure}

The lTRC approach affords insight into the origin of the phase duration effects of perturbation.
To capture these effects, we plot the integrand in \eqref{eq:T1_shifting_boundary} as well as the accumulated integral up to time $t$ as functions of $t$. Specifically, for $t\in[t^\text{in},t^\text{out}]$, the integrand depends on the three terms
\begin{equation}
\label{eq:integrand}
\left\{\begin{aligned}
&\eta(x^\text{in})\cdot\frac{\partial x_{\mu}^\text{in}}{\partial\mu}\bigg|_{\mu=0},\quad&t=t^\text{in}\\
-&\eta(x^\text{out})\cdot\frac{\partial x_{\mu}^\text{out}}{\partial\mu}\bigg|_{\mu=0},\quad&t=t^\text{out}\\
&\eta(\gamma(t))\cdot\frac{\partial F_{\mu}(\gamma(t))}{\partial\mu}\bigg|_{\mu=0},\quad&t^\text{in}<t<t^\text{out}.
\end{aligned}\right.
\end{equation}
Since the exit boundaries are fixed, $-\eta(x^\text{out})\cdot\frac{\partial x_{\mu}^\text{out}}{\partial\mu}\big|_{\mu=0}$ is  zero in this scenario.
Figure~\ref{fig:lTRC_IR_SR}(a) shows the timing responses to $d_1$ perturbation that occur in the three active phases over one period.
We observe for phase 1 (black trace) that the timing response just before the end of cell 1 activation is dominant while the effect of the perturbation on the entry to phase 1 is insignificant.
The only effect of the $d_1$ change on phase 2 (blue trace) and phase 3 (red trace) is the slight modulation of their jump-up points.
(We note that the local timing response curve in this setting has a qualitatively similar shape to the \emph{infinitesimal phase response curve} (iPRC) (cf.~\cite{brown2004phase,ermentrout2010,schwemmer2012theory}), and we discuss their relationship in Appendix~\ref{appssec:Releationship between iPRC and lTRC}.)

To qualitatively explain what the lTRC analysis reveals about the dominant effect of the jump-down of cell 1 on the  response to perturbation, we consider the fast-slow decomposition of the relaxation oscillator system, which makes use of the interplay of spatial and temporal measures of distance between trajectories \citep{terman1998,rubinhandbook}.
Recall that for the intrinsic release mechanism, the phase transition is triggered when the active cell reaches the appropriate right knee of the active $v$-nullcline. As $d_1$ increases and the right knee, or fold, of the $v_1$-nullcline shifts to a lower position, cell 1 spends more time in the active phase.
Specifically, as cell 1 progresses through the active state, it evolves toward the fixed point (the intersection of the free $v_1$-nullcline and the $h_1$-nullcline in Figure~\ref{fig:IR_phase1}), which lies just beyond the knee, until the knee is reached. Thus, the dynamics of cell 1 exponentially decelerates as the projected trajectory approaches the jump-down point, as shown by the contraction of points around the right knee in Figure~\ref{fig:IR_phase1}.
Hence, despite a small spatial difference for the $v_1$-nullcline caused by a slight change in $d_1$, the difference in time needed to cover that spatial spread, and hence the difference in the duration of active phase 1, is significant.

Note that as $d_1$ varies, the entry point of cell 1 into the active phase also changes (upper right corner of Figure~\ref{fig:IR_phase1}). A larger $d_1$ results in a smaller $h_1$ upon entry of cell 1 to the active phase,  and thus a shorter time spent in its active region.
This reduction partially counters the delay around the right knee.
However, since the voltage changes much more rapidly at the start of the active state than at the end, the spatial perturbation upon entry has negligible temporal effects.

Next, consider the effect of the $d_1$-perturbation on the active phase of cell 2, the next cell to activate after cell 1.
The change of $d_1$ does not affect the nullclines of $v_2$ or the jump-down point of cell 2 to the silent phase.
The timing of the entry of cell 2 into the active phase is controlled by the jump-down of cell 1 from the right knee of the $v_1$-nullcline, which releases cell 2 from inhibition.
When cell 1 spends more time before the end of its active state, cell 2 correspondingly spends more time in the silent state. However, since the position of cell 2 is close to the silent phase fixed point (with extremely low rate of change), the added time in the silent phase does not make much of a  difference in the spatial position at which cell 2 enters the active phase and hence its active phase duration.

\begin{figure}
\centering
\includegraphics[width=9cm]{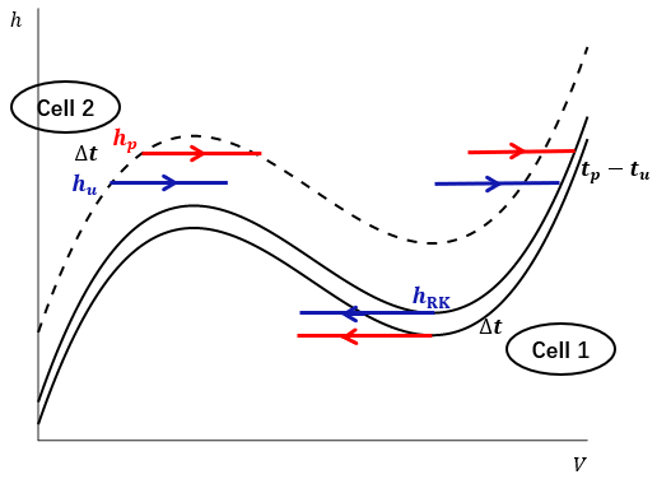}
\caption{\label{fig:time_compare} A schematic diagram for the intrinsic release mechanism, illustrating the jumps of cell 1 and cell 2 when $d_1$ is positively perturbed. Dashed black: inhibited $v$-nullcline for cell 2. Upper solid black: uninhibited $v$-nullcline for cell 2 and for cell 1 in the default $d_1$ case. Lower solid black: uninhibited $v_1$-nullcline for the perturbed $d_1$ case. Red (blue) traces represents the jumps when $d_1$ is perturbed (unperturbed). The extra time spent by cell 1 to reach the right knee $h_\text{RK}$ after perturbation is equivalent to the extra time that cell 2 spends in the silent phase, traveling from $h_u$ to $h_p$, denoted by $\Delta t$. When cell 2 enters the active phase, the time difference between the red and blue landing points on the uninhibited $v$-nullcline is much smaller than $\Delta t$; that is, $t_p-t_u\ll\Delta t,$ where $t_u$ (resp., $t_p$) represents the passage time for cell 2 from its arrival in the active phase until it reaches the right knee of the $v$-nullcline without (resp., with) the perturbation.}
\end{figure}

Specifically, Figure~\ref{fig:time_compare} shows a schematic illustration of the jump-up of cell 2 when $d_1$ is positively perturbed.
The longer time of cell 2 in the silent phase is equal to the extra time of cell 1 in the active phase, denoted by $\Delta t$ in the figure.
However, when cell 2 enters the active phase, the time difference from its perturbed to its unperturbed position is highly compressed. We can in fact show that
$$0<t_p-t_u\ll\Delta t,$$
where $t_u$ and $t_p$ represent the unperturbed time and perturbed time that cell 2 spends in the active phase, respectively.

Consider the extra journey of cell 2 during the silent phase in the perturbed case.
The evolution of the gating variable $h_2$ is approximated by
\begin{equation*}
    h_2'=(h_L-h_2)/\tau_L,
\end{equation*}
for some values of $h_L$ and $\tau_L$.
Hence, the extra time in the silent phase (from point $h_u$ to $h_p$ in Figure~\ref{fig:time_compare}) is given by
\begin{align}
\label{eq:Delta_t}
    h_p=h_L+(h_u-h_L)e^{-\Delta t/\tau_L}\nonumber\\
    \Longrightarrow\quad\Delta t=\tau_L\ln{\left(\frac{h_u-h_L}{h_p-h_L}\right)}.
\end{align}
On the right branch, $h_2$ approximately follows
\begin{equation*}
    h_2'=-h_2/\tau_R,
\end{equation*}
for some value of $\tau_R$.
Since the jump-down position of cell 2 is invariant with respect to the $d_1$ perturbation, the difference between the perturbed and unperturbed active durations of cell 2, which is also equal to the time of passage from the perturbed landing point of cell 2 in the active phase to its unperturbed landing point, can be computed from
\begin{align}
\label{eq:tp-tu}
    &\qquad\quad h_\text{RK}=h_pe^{- t_p/\tau_R}=h_ue^{-t_u/\tau_R}\nonumber\\
    &\Longrightarrow\quad t_p=\tau_R\ln{\left(\frac{h_p}{h_\text{RK}}\right)},\;t_u=\tau_R\ln{\left(\frac{h_u}{h_\text{RK}}\right)}\nonumber\\
    &\Longrightarrow\quad t_p-t_u=\tau_R\ln{\left(\frac{h_p}{h_u}\right)}.
\end{align}
To compare these quantities numerically, we choose $h_L=h_\infty(-62)$ and $\tau_L=\tau_h(-62)$, based on a typical value of $v_2$ in the silent phase, and we assume $\tau_R=\tau_h(-30)$ to give $t_p-t_u$ its largest possible value.
When $d_1$ is increased by $0.05$, we have $h_L=0.9751, \tau_L^{-1}=0.0321, \tau_R^{-1}=0.0137, h_p=0.8817, h_u=0.8814$; by \eqref{eq:Delta_t} and \eqref{eq:tp-tu} we obtain
$$t_p-t_u=0.0248,\qquad\Delta t=0.0999.$$
Note that because we chose the value of $\tau_R$ that occurs when the trajectory is near the right knee, which represents the largest value that $\tau_h(v_2)$ takes during the active phase, the difference $t_p-t_u$ is in fact much smaller than $0.0248$. This analysis confirms our temporal-spatial observation that the transition to the active state greatly compresses the temporal difference of cell 2 active phase durations (perturbed versus unperturbed), relative to the temporal difference of active phase durations for cell 1.
The jump down of cell 2 releases cell 3 to the active phase, so cell 3 inherits the insignificant active phase duration change from cell 2.
The fast-slow decomposition analysis above is in accordance with the lTRC characteristics (Figure~\ref{fig:lTRC_IR_SR}(a)).

\paragraph{Synaptic release}
Following a similar argument, we can uncover the sensitivity of the individual activation phases for the synaptic release mechanism.
Note that the two release mechanisms have the same nullcline structure (see Figures~\ref{fig:IR_solution}, and \ref{fig:SR_solution}) and we expect that very similar underlying mechanisms are at work in the two cases.

Table~\ref{tab:SR_duration_difference} lists the duration changes with reference to the symmetric case.
The duration of active phase 1 has a positive response to the increase of $d_1$ while the durations of phase 2 and phase 3 are insensitive.
Compare the sensitivity of cell 1 activation between the two release mechanisms (Tables~\ref{tab:IR_duration_difference},~\ref{tab:SR_duration_difference}).  While both mechanisms yield increases in the phase 1 duration,
$\Delta t_1$, the duration change for the synaptic release has a considerably smaller magnitude than $\Delta t_1$ for the intrinsic release.
This result is evident from the local timing response analysis as shown in Figure~\ref{fig:lTRC_IR_SR}.
The  plot of the integrand of $T_1$ in equation \eqref{eq:T1_shifting_boundary} indicates that although the integrand at the end of phase 1 grows to a large positive value under the synaptic release mechanism, the elevation in the synaptic release case occurs much more briefly,  resulting in a weaker temporal effect than results due to perturbation in the intrinsic mechanism.

\begin{table}
\centering
\caption{\label{tab:SR_duration_difference} Synaptic release: duration changes in three individual active phases, by direct numerical simulation or lTRC calculation. The durations in the default case ($d_1=d_2=d_3=1$) are $t_1=t_2=t_3=20.6558.$ The duration of phase 1 changes monotonically with $d_1$, and the durations of the other two phases are almost invariant.}
\begin{tabular}{|lcccclccc|}
\hline
Simulated difference&$\Delta t_1$&$\Delta t_2$&$\Delta t_3$&&lTRC difference&$\Delta t_1$&$\Delta t_2$&$\Delta t_3$\\
\hline
$\mu=0.05$&0.0245&-0.0002&0.0006&&$\mu=0.05$&0.0203&-0.0002&0.0006\\
\hline
$\mu=-0.05$&-0.0245&0.0002&-0.0007&&$\mu=-0.05$&-0.0203&0.0002&-0.0007\\
\hline
\end{tabular}
\end{table}

In the synaptic release mechanism, the synaptic threshold $\theta_\text{I}$ plays the key role in determining the transition of each cell from the active state to the silent state.
As shown in Figure~\ref{fig:SR_phase1}, as $d_1$ is adjusted, the voltage coordinate of the jump-down point of cell 1 remains almost equal to $\theta_\text{I}$ while the coordinate of the gating variable $h_1$ changes.
Because cell 1 only has to reach $v_1 = \theta_\text{I} > v_\text{RK}$ for the transition to begin, the smallest values of $\tau_h$ from the intrinsic-release case do not occur, and hence the difference in active phase duration for cell 1 is smaller in the synaptic release case.
This again translates into small differences in the active phase durations for cells 2 and 3.




\paragraph{Synaptic escape}
Unlike the release mechanisms, where the active cell itself has primary control of the transition, here the duration of the active phase is predominantly controlled by the silent cell.
Specifically, each transition is initiated when one of the silent cells hits the synaptic threshold, thus inhibiting the active cell and inducing a transition.
As $d_1$ increases, the inhibited $v_1$-nullcine shifts downwards,
and so does the escape position of the silent cell 1; the inhibited voltage nullclines, as well as the escape positions, of the other two cells are invariant.
In the following, we first show the lTRC characteristics associated with the synaptic escape mechanism.
Next, we provide a geometric argument to explain why the active phase duration of cell 1 increases while the active phase durations of cell 2 and cell 3 decrease when $d_1$ is increased.
Finally, we strengthen our argument by giving an analytical calculation to show that under the assumption that $t_1$ increases (which can be justified by the geometric argument), both of $t_2$ and $t_3$ must decrease.

\begin{figure}
\centering
\includegraphics[width=10cm]{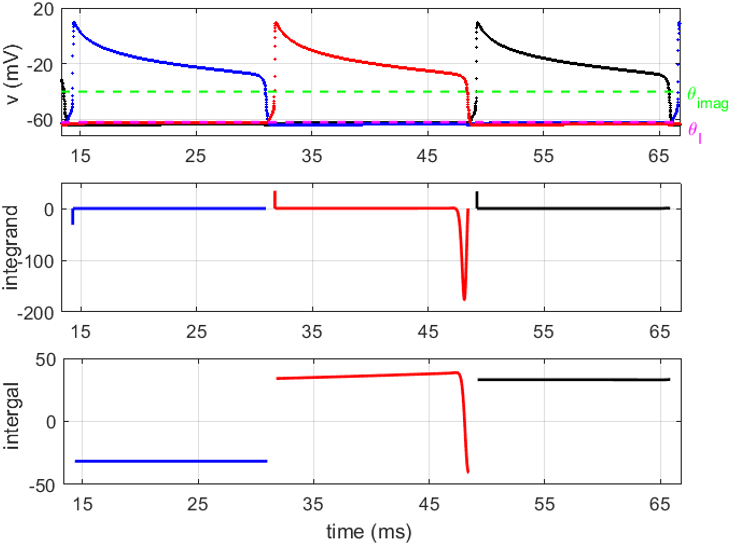}
\caption{\label{fig:lTRC_SE} Timing responses for the synaptic escape mechanism, measured by equation \eqref{eq:T1_shifting_boundary}.
Colors are as in Figure~\ref{fig:lTRC_IR_SR}.
Here we adopt an artificial synaptic threshold $\theta_\text{imag}$ to demarcate the start and end of the active phases.
When $d_1$ is increased (by perturbation $\mu=0.01$), all three of the active phases have significant responses: cell 2 (blue) exhibits an initial negative deflection; cell 3 (red) starts with a positive response that is subsequently overcome by a more significant negative peak; cell 1 (black) shows a positive deflection when it enters the active phase.}
\end{figure}

Table~\ref{tab:SE_duration_difference} shows the changes in the three active phase durations when a static perturbation $\mu=\pm 0.01$ is applied to $d_1$.
The first-order changes in the phase durations given by the lTRC formula \eqref{eq:T1_shifting_boundary} match well with direct numerical simulation.
The timing response characteristics of the synaptic escape mechanism are shown in Figure~\ref{fig:lTRC_SE}.
As $d_1$ increases, the silent cell 1 jumps up earlier, as indicated by the initial positive deflection of the black curve in the integrand plot.
The earlier jump-up of cell 1 results in a lower position of the  double-inhibition notch of cell 2 in its silent phase and a higher jump-down position of active cell 3, as shown in Figure~\ref{fig:SE_sketch_123} (see below for a more detailed discussion of Figure~\ref{fig:SE_sketch_123}).
These further lead to a later jump-up of cell 2 and an earlier jump-up of cell 3, which are in accordance with the initial negative deflection of the blue curve and positive deflection of the red curve in the integrand plot of Figure~\ref{fig:lTRC_SE}.
Right before the end of the active phase of cell 3, a significant negative response occurs and counters the initial positive response.
This negative response is due to the early escape of cell 1, which advances the termination of the active phase of cell 3.

\begin{table}
\centering
\caption{\label{tab:SE_duration_difference} Synaptic escape:  changes in durations of the three active phases, obtained by direct numerical simulation or lTRC calculation. The durations in the default case ($d_1=d_2=d_3=1$) are $t_1=t_2=t_3\approx16.6590.$ A one percent positive perturbation to $d_1$ induces large deviations to all three phases, with a positive response of $t_1$ and a negative response of both $t_2$ and $t_3$.}
\begin{tabular}{|lcccclccc|}
\hline
Simulated difference&$\Delta t_1$&$\Delta t_2$&$\Delta t_3$&&lTRC difference&$\Delta t_1$&$\Delta t_2$&$\Delta t_3$\\
\hline
$\mu=0.01$&0.3269&-0.3198&-0.3978&&$\mu=0.01$&0.3323 &-0.3160&-0.4068\\
\hline
$\mu=-0.01$&-0.3412&0.3291&0.4079&&$\mu=-0.01$&-0.3373&0.3350&0.4005\\
\hline
\end{tabular}
\end{table}

\begin{figure}
\centering
\includegraphics[width=16.5cm]{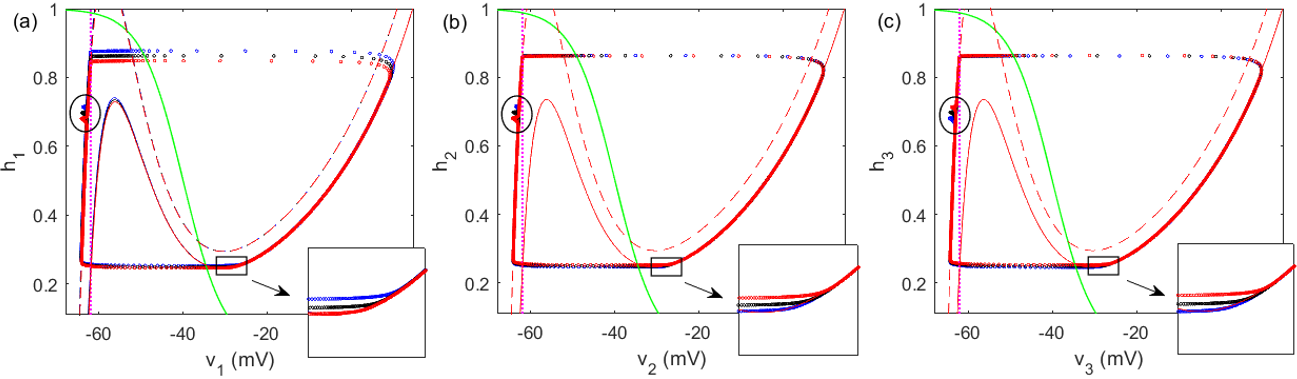}
\caption{\label{fig:SE_phase123} Synaptic escape mechanism: numerically generated nullclines and trajectories in the phase plane of \textbf{(a)} cell 1, \textbf{(b)} cell 2, and \textbf{(c)} cell 3, for the three different values $d_1=0.97$ (blue), $d_1=1$ (black) and $d_1=1.03$ (red).
Because the $d_1$-perturbation only influences the $v_1$-nullclines, the jump-up position of cell 1 is lowered as $d_1$ increases and the jump-up positions of cell 2 and cell 3 are unchanged.
Small changes occur in the jump-down positions of all three cells (rectangles and insets).
The ``double-inhibition notch" in the circles play an important role in determining the active phase durations, as explained in the text and Figure~\ref{fig:SE_sketch_123}.}
\end{figure}

\begin{figure}
\centering
\includegraphics[width=15.5cm]{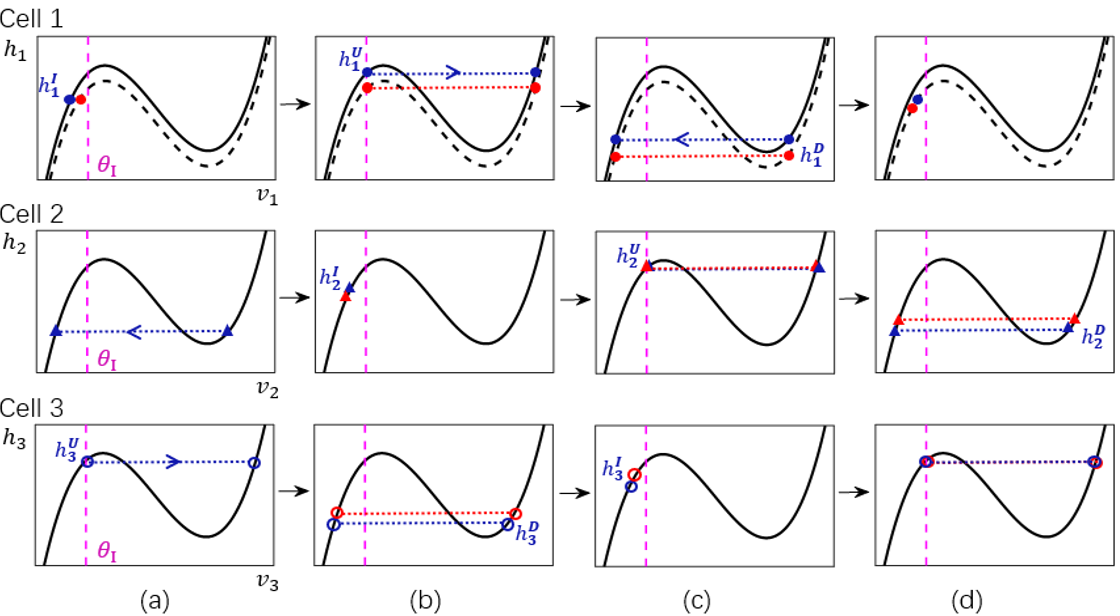}
\caption{\label{fig:SE_sketch_123} A schematic diagram for the synaptic escape mechanism illustrating locations of the three cells as time evolves (a)$\rightarrow$(b)$\rightarrow$(c)$\rightarrow$(d).
Suppose a positive perturbation is applied to $d_1$ when cell 3 is at the escape position, which immediately results in the position change of cell 1 (panels (a)).
For visual convenience, we show only one $v_i$-nullcline per input level for cell $i$.
The $v_1$-nullcline  is lowered by the positive perturbation (dashed cubic curve in the top panels), while the $v_2$-nullcline and $v_3$-nullcline are invariant.
Magenta lines indicate the synaptic threshold $\theta_\text{I}$.
Notations $h_i^U$ and $h_i^D$ represent the $h_i$-coordinate of cell $i$ at the jump-up position and jump-down position, respectively;
notation $h_i^I$ on the left branch of the nullcline represents the location of the ``double-inhibition notch" of silent cell $i$ when some other cell is at escape.
Red marks indicate the new locations of the cells subjected to the increased $d_1$, and blue marks indicate the corresponding old locations with the default $d_1$.
Note that in the cell 1 panel of (d), we use the blue dot to mark the location of the red dot from panel (a) that resulted from the initial perturbation of $d_1$.
See text for analysis of how the locations of the crucial points and the active phase durations change.
The corresponding numerical simulation figures are shown in Figure~\ref{fig:SE_phase123}.}
\end{figure}

Figure~\ref{fig:SE_phase123} shows the phase portraits and trajectories obtained by numerical simulation for the synaptic escape mechanism.
Note that there is a notch circled in each cell panel (see also Figure~\ref{fig:IE_solution_symmetric}), arising from double inhibition from the other two cells (which plays an important role in changing the active phase durations).
Figure~\ref{fig:SE_sketch_123} provides a schematic illustration of the phase portraits for the three cells.
For visual convenience, we just show one $v$-nullcline per input level for each cell (i.e., we ignore the shift in $v$-nullclines due to inhibition).
One can also refer to Figure~\ref{fig:SE_phase123} for the phase portraits and trajectories obtained by numerical simulation in this case.
Red marks represent the locations of the cells resulting from an increase in $d_1$ and blue marks represent the corresponding locations with the default $d_1$.
Suppose $d_1$ is perturbed at the moment when cell 3 is at the jump-up position $h_3^U$ (see panels (a)).
Note that the jump-up of cell 3 causes cell 2 to jump down (from the right to the left branch of its $v$-nullcline).
Since the $v_1$-nullcline shifts downwards and the current position of cell 1, at the double-inhibition notch, denoted by $h_1^I$, moves to the right,  the passage time of cell 1 to its escape position $h_1^U$ (panels (b)) is shortened. Correspondingly, the active duration of cell 3, $t_3$, 
is shortened (or equivalently, the jump-down position of cell 3, $h_3^D$, is higher), and the double-inhibition notch location of cell 2 on the inhibited $v_2$-nullcline, denoted by $h_2^I$, is lower (panels (b)).
In panels (c), since the escape position of cell 2, $h_2^U$, is invariant with respect to $d_1$, the passage time of cell 2 to from $h_2^I$ to $h_2^U$ becomes longer, which results in a longer $t_1$, a lower $h_1^D$, and a higher $h_3^I$.
Such higher $h_3^I$ leads to a shorter passage time of cell 3 to its escape position, and thus, $t_2$ is shortened, $h_2^D$ is higher, and $h_1^I$ is lower (see panels (d)).
Note that in the cell 1 panel of (d), we use the blue dot to mark the location of the red dot from panel (a) that resulted from the initial perturbation of $d_1$.
Since $h_1^I$ is lower at the end of the cycle, in panel (d), than it was initially, if we continue to track the cells following a similar argument, then we see that the active $t_3$ in the second iterate is no longer quite as short as it was in the first iterate.

To represent this situation analytically, we will introduce the notation $t_i^A$ to denote active phase durations (previously $t_i$) and $t_i^S$ to denote silent phase durations.  
Let $t_i^A(k)$ denote the active phase duration of cell $i$ on iterate $k$, where $k=0$ corresponds to the unperturbed solution.
From above, we have 
\begin{equation}
\label{eq:t3_first_second}
t_3^A(2)>t_3^A(1).
\end{equation}
We aim to show that the unperturbed active duration of phase 3 is longer than the perturbed active duration of phase 3. 
That is, 
\begin{equation*}
t_3^A(0)> t_3^A(k),\quad k\geq1.
\end{equation*}
Since the active duration of cell 3 is equal to the passage time of cell 2 from point $h_2^D$ to $h_2^I$ in the silent phase, it is sufficient to consider the $h_2$-dynamics of cell 2.
We approximate this dynamics with the ODE 
$$h_2'=(h_L-h_2)/\tau_L$$
for some values of $h_L$ and $\tau_L$ since $h_{\infty}, \tau_h$ are approximately constant over the relevant range of voltages. By solving the ODE, we obtain at iterate $k$
\begin{align*}
t_3^A(k)=\tau_L\ln{\left(\frac{h_L-h_2^D(k)}{h_L-h_2^I(k)}\right)}.
\end{align*}
Comparing $k=0$ and $k=2$ yields
\begin{align*}
 t_3^A(2)-t_3^A(0)=\tau_L \ln{\left(\frac{h_L-h_2^D(2)}{h_L-h_2^D(0)}\cdot\frac{h_L-h_2^I(0)}{h_L-h_2^I(2)}\right)}.
\end{align*}
Note that when cell 2 is silent, the $h_\infty(v_2)$ curve always lies above $h_2$, no matter what value of $h_L$ is taken to approximate it.
This indicates that $h_L-h_2^D>h_L-h_2^I>0$.
Combining this observation with the discussion in the previous paragraph gives
$$h_2^I(2)<h_2^I(0)<h_L,\qquad h_2^D(0)<h_2^D(2)<h_L.$$
Therefore we have $t_3^A(0)>t_3^A(2)$. 
Together with \eqref{eq:t3_first_second}, the active phase durations for cell 3 in the first two iterations obey the following rule
$$t_3^A(0)>t_3^A(2)>t_3^A(1).$$
Repeating similar arguments, we obtain
\begin{align*}
    &t_3^A(0)>t_3^A(2)>t_3^A(1),\\
    &t_1^A(0)<t_1^A(2)<t_1^A(1),\\
    &t_2^A(0)>t_2^A(2)>t_2^A(1).
\end{align*}
Continuing in this way, we get a sequence of active phase times, which shows that the durations $t_1^A(k)$ will stay above the unperturbed $t_1^A(0)$ while $t_2^A(k), t_3^A(k)$ will stay below $t_2^A(0), t_3^A(0)$, respectively.
We can conclude that when subject to a static positive $d_1$ perturbation, $t_1^A$ increases while $t_2^A$ and $t_3^A$ decrease.

The decrease of $t_2^A$ and $t_3^A$ can also be obtained by a direct calculation.
Consider the dynamics of cell 1, with active (silent) duration $t_1^A$ ($t_1^S$) and with $h_1$-coordinate ranging from $h_1^U$ to $h_1^D$.
Assume that $t_1^A$ increases when $d_1$ is positively perturbed.
As shown in Figure \ref{fig:SE_sketch_123}, the perturbation lowers $h_1^U$ and thus lowers $h_1^D$ as well.
In the silent phase, the time of cell 1 from $h_1^D$ to $h_1^U$ is $t_1^S=t_2^A+t_3^A$, and thus, by approximating the ODE for $h_1$, we have
\begin{align*}
    &h_1'=(h_L-h_1)/\tau_L\\
    \Longrightarrow\quad& h_1^U=h_1^De^{-(t_2^A+t_3^A)/\tau_L}+h_L(1-e^{-(t_2^A+t_3^A)/\tau_L}).
\end{align*}
Denote $x(d_1)=e^{-(t_2^A+t_3^A)/\tau_L}$ to obtain
\begin{align*}
   h_1^U(d_1)&=h_1^D(d_1)x(d_1)+h_L(1-x(d_1))\\
   \Longrightarrow\quad x(d_1)&=\frac{h_L-h_1^U(d_1)}{h_L-h_1^D(d_1)}.
\end{align*}
Due to the compression around the right knee of the active $v_1$-nullcine, the magnitude of the decrease in $h_1^U(d_1)$ is larger than the magnitude of the decrease in $h_1^D(d_1)$.
Hence, $x(d_1)$ increases, and $(t_2^A+t_3^A)$ decreases.

Now, suppose that $t_3^A$ increases; correspondingly, $t_2^A$ must decrease.
On the one hand, this indicates that $t_2^S=t_1^A+t_3^A$ increases.
On the other hand, since $t_2^A$ is shortened and $h_2^U$ is independent of $d_1$, then $h_2^D$ must be higher.
Therefore, $t_2^S$ must decrease, which is a contradiction.
Hence, $t_3^A$ decreases. A similar argument shows that $t_2^A$ also decreases.
In this way, we conclude that when the active phaseduration for cell 1 increases, the active phase durations for both cell 2 and cell 3 must decrease.
This geometric analysis and direct mathematical calculation show agreement with the timing response analysis shown in Figure~\ref{fig:lTRC_SE}.

\begin{figure}
\centering
\includegraphics[width=15.5cm]{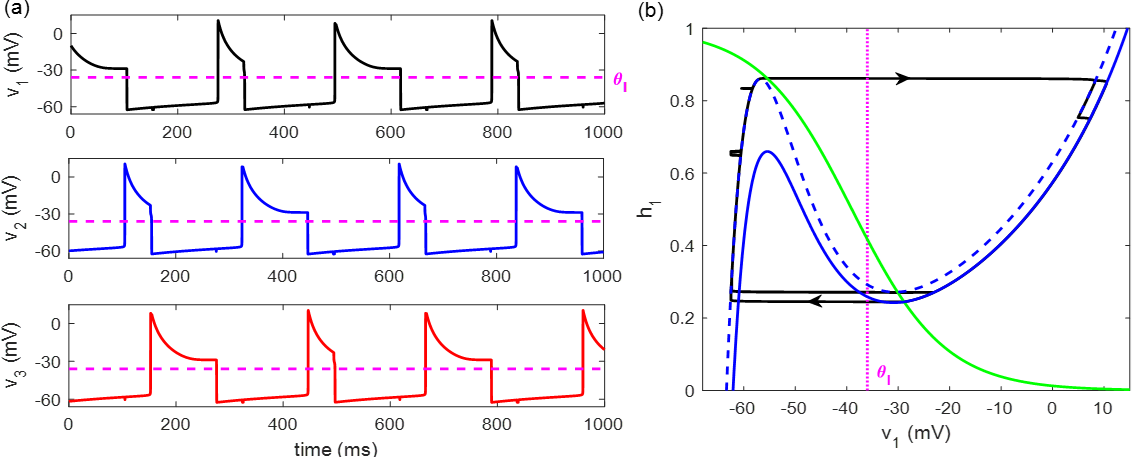}%
\caption{\label{fig:IE_solution_asymmetric} A double-period solution associated with the intrinsic escape mechanism. Colors and model parameters as in Figure~\ref{fig:IE_solution_symmetric}. The system exhibits a stable short-long alternation and two double-inhibition notches, with the nontrivial Floquet multipliers less than one in magnitude.}
\end{figure}

\paragraph{Intrinsic escape}
The system with phase transitions via the intrinsic escape mechanism has an  unstable equal-duration firing pattern (Figure~\ref{fig:IE_solution_symmetric}) and multiple stable short-long patterns (e.g., Figure~\ref{fig:IE_solution_asymmetric}, generated with the same parameter set as Figure~\ref{fig:IE_solution_symmetric}).
Numerically we find that the equal-duration solution in Figure~\ref{fig:IE_solution_symmetric} has two unit Floquet multipliers, while for the short-long solution in Figure~\ref{fig:IE_solution_asymmetric}, all Floquet multipliers except the trivial Floquet multiplier are strictly smaller than one in modulus.
Due to the particularity of this mechanism, we do not give its sensitivity analysis.
Appendix~\ref{app:IE_instability} presents analysis of the instability of the equal-duration periodic solution for the intrinsic escape mechanism in the singular limit $\epsilon\rightarrow0$.

\subsection{Heteroclinic cycling model}
\label{ssec:sensitivity_HC}

\begin{figure}
\centering
\includegraphics[width=15cm]{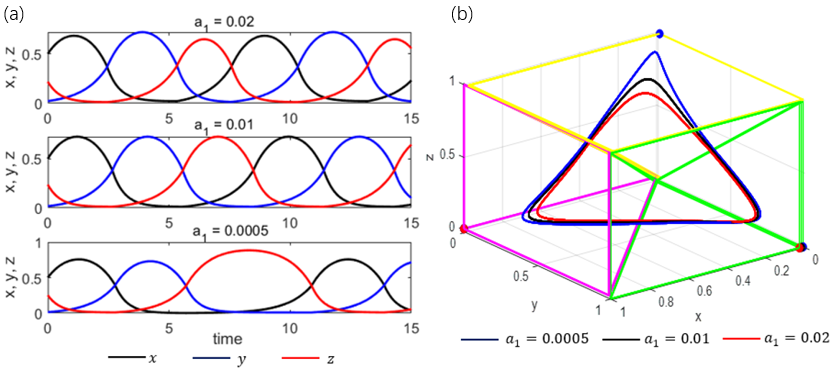}%
\caption{\label{fig:Aplysia_change_a1} Trajectories of the heteroclinic cycling model \eqref{eq:Aplysia_equations} at three values of $a_1$, with $a_2=a_3=0.01$.
\textbf{(a)}: time courses of pool $x$ (black), $y$ (blue) and $z$ (red) when $a_1=0.02$ (top), $a_1=0.01$ (middle) and $a_1=0.0005$ (bottom).
\textbf{(b)}: Trajectories in the 3D cubic domain, with $a_1=0.02$ (red), $a_1=0.01$ (black) and $a_1=0.0005$ (blue).
The active duration of pool $z$ shrinks most when $a_1$ increases, while that of pool $y$ shrinks least.}
\end{figure}

In this section, to study the controllability of the heteroclinic cycling model \eqref{eq:Aplysia_equations}, we apply a small sustained perturbation $\mu$ to $a_1=0.01$ (i.e., $a_1\rightarrow a_1+\mu$), with the other two control parameters $a_2$ and $a_3$ fixed at the default value of $0.01$.
As a reference, Figure~\ref{fig:Aplysia_change_a1} shows how the dynamics of the three pools evolves when $a_1=0.0005,\,0.01$, or $0.02$.
In what follows, we explain the sensitivity of each phase duration using the lTRC as well as direct analysis based on the vector field.

\begin{figure}
\centering
\includegraphics[width=9.5cm]{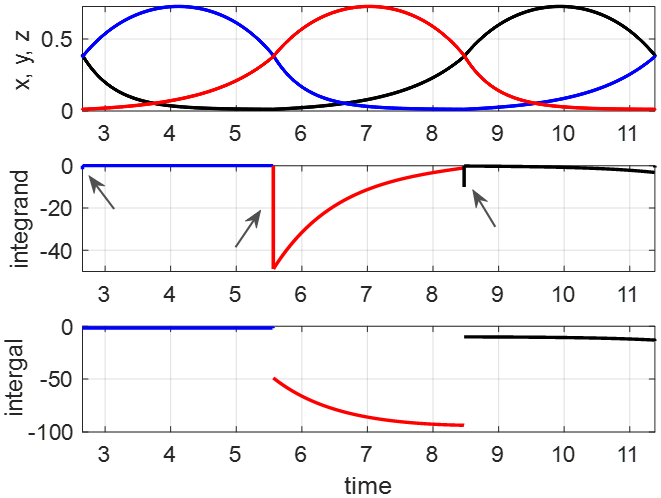}%
\caption{\label{fig:lTRC_Aplysia} Timing responses in the heteroclinic cycling model \eqref{eq:Aplysia_equations}.
Black, blue and red traces represent pools $x$, $y$ and $z$, respectively.
When a positive perturbation is applied to $a_1$, pool $z$ has a significant negative timing response throughout its active phase.
The initial displacement at the beginning of each phase (grey arrows) indicates the effect of the perturbation on the entry point to each phase (given by term A in equation \eqref{eq:T1_shifting_boundary}).}
\end{figure}

Figure~\ref{fig:lTRC_Aplysia} shows the timing responses of the heteroclinic cycling system computed from the lTRC ias given in equation \eqref{eq:T1_shifting_boundary}.
Phase 3 (red) has a significant negative response to the increase of $a_1$; phase 2 (blue) is almost invariant; phase 1 (black) has a relatively small negative response.
To understand the effect upon trajectory entry to each phase, consider the terms comprising ``term A" in equation \eqref{eq:T1_shifting_boundary}, which captures the effects of shifting the entry point to a given region, namely
$\eta(x^\text{in})\cdot\frac{\partial x_\mu^\text{in}}{\partial\mu}\big|_{\mu=0}$.
The timing response curve vector at the entry point to phase 3 is $\eta^\text{\uppercase\expandafter{\romannumeral3}}(x^\text{in})=(-48.89,\,0,\,0.06)^T$,
which shows that the first component should dominate the response.
The lTRC at the entry points to the other phases is given by cyclic permutation, so the contribution to the response at the entry to phase 2 is dominated by the third component, and the response upon entry to phase 1 is driven by the second component.
The timing sensitivity depends on the inner product between the lTRC and the derivative of the entry location  with respect to $\mu$.
The lTRCs are symmetric (under threefold rotation); the derivatives of the entry location vary from phase to phase.
Specifically, the entry location on phase 3 varies most, while the entry location on phase 2 varies least (as a reference, see Figure~\ref{fig:Aplysia_change_a1}(b)).
This accounts for the smaller or larger initial displacement effects seen in Figure \ref{fig:lTRC_Aplysia} (grey arrows).
Table~\ref{tab:Aplysia_duration_difference} lists the duration changes in the three individual phases, obtained either from direct numerical simulation or from the lTRC formula \eqref{eq:T1_shifting_boundary}, when a perturbation of size $|\mu|=0.0005$ is applied to $a_1=0.01$.

\begin{table}
\centering
\caption{\label{tab:Aplysia_duration_difference} Heteroclinic cycling model: duration changes in three individual phases, from direct numerical simulation or lTRC calculation. The durations in the default case ($a_1=a_2=a_3=0.01$) are $t_1=t_2=t_3=2.9080$. With a small positive perturbation to $a_1$, pool $z$ shows a significant decrease in the active duration ($\Delta t_3$); pool $y$ is insensitive ($\Delta t_2$); pool $x$ has a relatively small negative response ($\Delta t_1$).}
\begin{tabular}{|lcccclccc|}
\hline
Simulated difference&$\Delta t_1$&$\Delta t_2$&$\Delta t_3$&&lTRC difference&$\Delta t_1$&$\Delta t_2$&$\Delta t_3$\\
\hline
$\mu=0.0005$&-0.0070&-0.0010&-0.0460&&$\mu=0.0005$&-0.0070&-0.0010&-0.0471\\
\hline
$\mu=-0.0005$&0.0070&0.0010&0.0480&&$\mu=-0.0005$&0.0070&0.0010&0.0468\\
\hline
\end{tabular}
\end{table}

The vector field of the model is given in \eqref{eq:Aplysia_equations}.
During phase 1 (in which pool $x$ is active), a positive perturbation to $a_1$ only influences the evolution of $x$ and the boundary surfaces of phase 1.
When $dx/dt>0$,
the increase of $a_1$ decreases the speed of $x$, making $x$ grow slower; when $dx/dt<0$,
the increase of $a_1$ accelerates the decline of $x$.
Note that the exit surface of $x$ from phase 1 is $x=y+\frac{a_1+a_2}{2}$ (and the trajectory stays in phase 1 when $x\geq y+\frac{a_1+a_2}{2}$).
Compared with the default case, the effects of increasing $a_1$ on $dx/dt$ cause the $x$ on the left hand side of this condition to shrink, while the right hand side increases with $a_1$.

With perturbation $a_1\rightarrow a_1+\mu$, denote the perturbed $(x,y,z)$-solution and the unperturbed $(x,y,z)$-solution to be $(x_p(t),y_p(t),z_p(t))$ and $(x_u(t),y_u(t),z_u(t))$, respectively.
Let $(x_p^0,y_p^0,z_p^0)$ and $(x_u^0,y_u^0,z_u^0)$ represent the entry points for the perturbed case and unperturbed case, respectively.
Let $t_1$ be the unperturbed exit time from phase 1 and $t_1^*=t_1+\Delta t_1$ be the perturbed exit time; from the above observations, $\Delta t_1<0$.
We analytically derive that the first-order approximation for the duration change in phase 1 can be estimated by (see  Appendix~\ref{app:analytical_duration_heteroclinic_cycling} for the derivation)
\begin{align}
\label{eq:calculate_t1}
    \Delta t_1=\frac{\mu[1/2+(1-e^{-t_1})\rho]-e^{-t_1}(x^0_p-x_u^0)-(e^{-t_1}-e^{t_1})(y_p^0-y_u^0)\rho/2+e^{t_1}(y_p^0-y_u^0)}{1-(2+\rho)x_u(t_1)-\mu e^{-t_1}\rho-e^{-t_1}(x^0_p-x_u^0)-(e^{-t_1}+e^{t_1})(y_p^0-y_u^0)\rho/2-e^{t_1}(y_p^0-y_u^0)}+O(\Delta t_1^2)
\end{align}
When $\mu=0.0005$, from a direct numerical simulation, we have
\begin{align*}
    &t_1 = 2.9080,\quad x_u(t_1) = 0.3874,\quad x_u^0= 0.3773,\\ &x_p^0=0.3769,\quad y_u^0=0.0111,\quad y_p^0=0.0112,
\end{align*}
and \eqref{eq:calculate_t1} gives $\Delta t_1\approx-0.0071$.
Compared with Table~\ref{tab:Aplysia_duration_difference},
this result confirms that equation \eqref{eq:calculate_t1} provides an accurate estimate to the change of phase 1 duration.
We note that based on numerical calculations, the fidelity of this estimate depends on the inclusion of the impact of the perturbation on the positions of the entry points into the phase (i.e., $x_p^0,x_u^0,y_p^0,y_u^0$), which is consistent with the initial negative deflection of the black curve in Figure~\ref{fig:lTRC_Aplysia} (grey arrow).

For phase 2, in which pool $y$ is active, the entry points for different values of $a_1$ are extremely close (see Figure~\ref{fig:Aplysia_change_a1} when a very large perturbation is applied).
The change in $a_1$ influences neither $dy/dt$ nor $dz/dt$, which also do not depend on $x$.
Further, the exit surface, given by $y=z+\frac{a_2+a_3}{2}$, is also independent of $a_1$.
Therefore, the duration of phase 2 is insensitive to the $a_1$-perturbation.

When $z$ is active in phase 3, the increase of $a_1$ leads to a faster increase of $x$, which then results in a faster decay of $z$.
The exit surface is $z=x+\frac{a_1+a_3}{2}$ and the trajectory stays in phase 3 when $z>x+\frac{a_1+a_3}{2}$.
Hence, a decrease in $z$ on the left hand side of this condition and an increase in the right hand side due to both $x$ and $a_1$ indicate that the trajectory leaves phase 3 earlier than in the default case.
Similar to phase 1, we obtain
\begin{scriptsize}
\begin{align}
\label{eq:calculate_t3}
    \Delta t_3=\frac{\mu(1/2+\rho)+e^{-t_3}(z_p^0-z_u^0)+e^{-t_3}(x_p^0-x_u^0-\mu)\rho/2-e^{t_3}(x_p^0-x_u^0+\mu)(1+\rho/2)}{a_3\rho-1+(2+\rho)x_u(t_3)+e^{-t_3}(z_p^0-z_u^0)+e^{-t_3}(x_p^0-x_u^0-\mu)\rho/2+e^{t_3}(x_p^0-x_u^0+\mu)(1+\rho/2)+\frac{3a_1+a_3}{2}}+O(\Delta t_3^2).
\end{align}
\end{scriptsize}
From direct numerical simulation,
\begin{align*}
    &t_3=2.9080,\quad x_u(t_3)=0.3769,\quad x_u^0=0.0111,\\
    &x_p^0=0.0116,\quad z_u^0=0.3774,\quad z_p^0=0.3772,
\end{align*}
and by \eqref{eq:calculate_t3}, $\Delta t_3\approx-0.0448.$
Consistent with the direct numerical simulation and lTRC results in Table~\ref{tab:Aplysia_duration_difference}, equation \eqref{eq:calculate_t3} gives a good estimate of the phase 3 duration change.
Numerical calculations suggest that similarly to phase 1, the perturbation of the entry position to phase 3 has a significant influence on the phase 3 duration (as suggested by the large initial displacement of the red curve in Figure~\ref{fig:lTRC_Aplysia} (grey arrow)).

Note that such vector field analysis is of limited general utility because it requires the solutions of the model ODEs.
Since the results of this analysis give a good agreement with the timing response calculation, however, this example provides a reassuring demonstration of the relevance  of the lTRC analytical formula~\eqref{eq:T1_shifting_boundary}, which can be used to compute responses to perturbations in general cases.

\subsection{Competitive threshold-linear network}

\begin{figure}
\centering
\includegraphics[width=16cm]{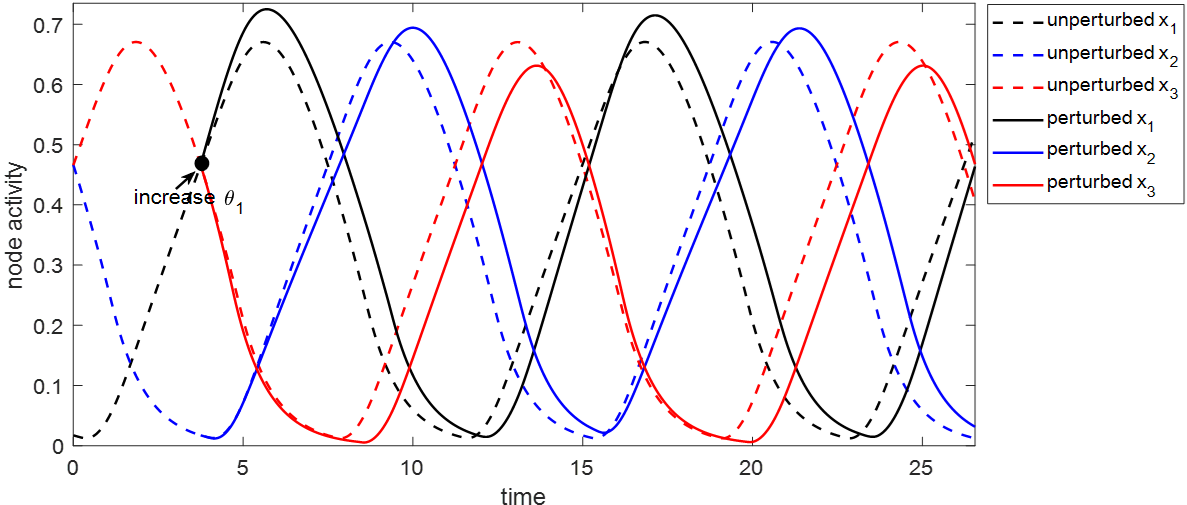}%
\caption{\label{fig:TL_compare_pert_unpert} Comparison of time courses for the competitive threshold-linear model in the unperturbed case and perturbed case.
Dashed traces: trajectory in the unperturbed case where $\theta_1=\theta_2=\theta_3=1$.
Solid traces: trajectory in the perturbed case where $\theta_1$ is increased to 1.05 at time 3.75 (solid black dot) when $x_1$ becomes active.
The increase of $\theta_1$ leads to a prolonged duration of $x_1$ activation and $x_2$ activation, in exchange for a shortened duration of $x_3$ activation.}
\end{figure}

In the competitive threshold-linear system \eqref{eq:threshold_linear_equations}, we impose a small static perturbation $\mu$ on the drive term $\theta_1$ to the first node $x_1$.
As an illustration, Figure~\ref{fig:TL_compare_pert_unpert} shows the trajectories in the default case where $\theta_1=\theta_2=\theta_3=1$ and in the perturbed case where $\theta_1$ is positively perturbed at the beginning of the first active phase.
(Note that for convenience of intuitive understanding, the perturbation here is applied since the first active phase, but in the following calculation we still impose a static perturbation along the limit cycle.)
We can understand the effects of perturbation as shown in Figure~\ref{fig:TL_compare_pert_unpert} by referring to equation (\ref{eq:threshold_linear_equations}).  The increase in $\theta_1$ implies that $W_{12}x_2 + W_{13}x_3$ must become more negative (recall that $W_{12}, W_{13}<0$) in order to cause $\dot{x}_1=0$ than was needed in the unperturbed case.  Although $x_2$ begins to increase while $x_1$ is still active, even if $x_2$ increased at its unperturbed rate, it would take longer for $x_2$ to grow enough for $W_{12}x_2 + W_{13}x_3$ to reach this more negative value; moreover, $x_2$ actually becomes slowed once $x_1$ exceeds its unperturbed maximum.  The longer time until $x_1$ starts decreasing and the larger maximal amplitude of $x_1$ translate into a longer active phase for cell 1.  Meanwhile, again with reference to equation (\ref{eq:threshold_linear_equations}) with $W_{31}=-1-\delta$, we see that the larger amplitude of $x_1$ makes $\dot{x}_3$ stay negative for a longer time than before and causes $x_3$ to reach a lower minimum value.
From this lower minimum, $x_3$ inhibits $x_2$ only weakly, allowing $x_2$ to stay larger than before, and it takes longer for $x_3$ to overtake $x_2$ than in the unperturbed case, resulting in a longer active phase for cell 2.
Meanwhile, $\dot{x}_1$ becomes positive earlier relative to the active phase of $x_3$ than before, due to the increase in $\theta_1$.  The resulting earlier increase in $x_1$  causes an earlier sign switch of $\dot{x}_3$ at a smaller $x_3$ value, resulting in a shorter active phase for cell 3.

\begin{figure}
\centering
\includegraphics[width=9.5cm]{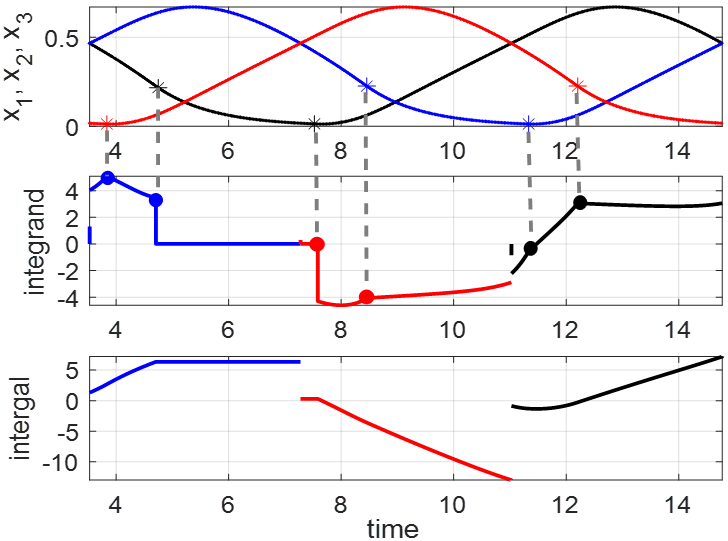}
\caption{\label{fig:lTRC_TL} Timing responses for the competitive threshold-linear model \eqref{eq:threshold_linear_equations}.
Black, blue and red traces represent nodes $x_1$, $x_2$ and $x_3$, respectively.
Asterisks in the top panel mark nonsmooth points of the  solution trajectory (see also Figure~\ref{fig:TL_reference}).
Correspondingly, the integrand of the lTRC in the middle panel inherits the nonsmooth dynamics (solid dots).}
\end{figure}

Using the lTRC calculation, Table~\ref{tab:TL_duration_difference} lists the duration change for each individual active phase in response to the perturbation of magnitude $|\mu|=0.01$ on $\theta_1$, and Figure~\ref{fig:lTRC_TL} shows the local timing responses.
Consistent with direct analysis above, all three phases are sensitive to the perturbation: the first two phases are prolonged under a positive perturbation, while the third phase is shortened.

\begin{table}
\centering
\caption{\label{tab:TL_duration_difference} Competitive threshold-linear model: active phase duration changes in three individual phases, from direct numerical simulation or lTRC calculation. The durations in the default case ($\theta_1=\theta_2=\theta_3=1$) are $t_1=t_2=t_3=3.7470$. With a small positive perturbation to $\theta_1$, nodes $x_1$ and $x_2$ display a positive response in their active duration; node $x_3$ displays a large negative response. With a small negative perturbation to $\theta_1$, the opposite responses occur.}
\begin{tabular}{|lcccclccc|}
\hline
Simulated difference&$\Delta t_1$&$\Delta t_2$&$\Delta t_3$&&lTRC difference&$\Delta t_1$&$\Delta t_2$&$\Delta t_3$\\
\hline
$\mu=0.01$&0.0730&0.0640&-0.1290&&$\mu=0.01$&0.0721&0.0636&-0.1298\\
\hline
$\mu=-0.01$&-0.0670&-0.0640&0.1350&&$\mu=-0.01$&-0.0689&-0.0636&0.1377\\
\hline
\end{tabular}
\end{table}

We note that the phase duration responses are quite different from those in the heteroclinic cycling case (compare Table~\ref{tab:Aplysia_duration_difference}).
The heteroclinic cycling case allows independent phase modulation --- selectively increasing the active duration of $i$-th unit can be attained by decreasing the drive input to the ($i$+1)-st unit.
In contrast, in the competitive threshold-linear case, implementing perturbation on one phase gives rise to simultaneous changes in the other phases.
A key difference between these two cases is that the heteroclinic cycling model features the passage near a saddle point in each phase, which the threshold-linear model lacks.  Thus, the difference in responses to perturbations between these cases suggests that the slowing of trajectories near saddle points may make an important contribution to the timing responses to perturbation  in the heteroclinic cycling case.

\section{Discussion}
\label{sec:Discussion}



We have considered three different dynamical systems frameworks for the production of triphasic oscillations in which three components take turns becoming active in a fixed order.
In each framework, we have presented a computational and mathematical analysis of the effects induced by changes in a parameter, which represents the strength of tonic input to one component, on the active phase durations of all of the components in the network.
We find that distinct phase duration responses arise when phase transitions occur through different mechanisms.
Specifically, we arrive at the following observations.

\paragraph{Relaxation oscillator model}

Each of the four mechanisms underlying the relaxation oscillator model features its own phase modulation characteristics.
\begin{itemize}
  \item Intrinsic release: an increase in the excitatory gain factor $d_1$ prolongs the active phase of cell 1 and leaves those of cells 2 and 3 almost unchanged.
  The mechanism involves the fast-slow dynamics of the model. Although the increased $d_1$ advances the entry of cell 1 to the active phase, the small spatial change around the jump-down (exit) position of cell 1 from the active phase translates into a significant temporal extension. As cell 1 progresses through the active state, the dynamics exponentially decelerates towards the exit position.
  \item Synaptic release: similarly to intrinsic release, an increase in $d_1$ increases the duration of cell 1's active phase, but it is a quantitatively smaller effect in the synaptic release case than in the intrinsic release case.
  The mechanism by which increasing $d_1$ affects the synaptic release case is qualitatively similar to that of the intrinsic release case.
  The  smaller effect size is due to the different jump-down positions in the two release mechanisms; in the synaptic case, the jump-down of cell 1 is farther from the fixed point, with less slowing down, resulting in the less significant temporal increase.
  \item Synaptic escape: an increase in $d_1$ prolongs the active phase of cell 1 and shortens the active phase durations of cells 2 and 3.
  In contrast to the release mechanisms, in this case an increase in $d_1$ allows cell 1 to enter its active phase earlier in the cycle than it otherwise would, leading to several changes in the points at which the cells in the circuit initiate jumps between phases and to the compound effects on active phase durations of all three cells.
  \item Intrinsic escape: although the two-unit half-center oscillator (HCO) with intrinsic escape produces stable oscillations, the 3-cell circuit produces an unstable equal-duration pattern and multiple stable short-long alternating patterns.
  In this study, we focus on patterns with equal phase durations, and thus the  response sensitivity of the stable patterns in this case are outside of the scope of our analysis.
\end{itemize}
Overall, we conclude that in the release cases, the system can selectively increase the active duration of a single unit with minimal change in the active phases of the other units, just by increasing the drive to that specific unit.
However, in the synaptic escape case, an increase in the active duration of one unit is accompanied by non-negligible decreases in the active durations of the other units.
In other words, independent active phase modulation is possible with release transitions but not escape transitions.

\paragraph{Heteroclinic cycle model}
An increase in the excitatory neural drive $a_1$ leads to a slight decrease in the active phase duration of the first cell, a negligible change in the active phase duration of the second cell, and a large decrease in the active phase duration of the third cell.
The mechanism appears mainly to involve shifting the entry point of the trajectory to the domain governed by the vector field around the saddle point corresponding to elevated activity in the cell receiving the increased drive.
Nearly independent active phase modulation can be achieved in this model;
in contrast to the relaxation oscillator model,  selectively increasing the active duration of the $i$-th unit requires decreasing the input to the succeeding, ($i$+1)-st unit.

\paragraph{Competitive threshold-linear network model}
Increasing the external drive $\theta_1$ to unit 1 prolongs the duration of the active phases of both cell 1 and cell 2, while decreasing the active phase duration of cell 3.
This mechanism increases the active duration of the $i$-th unit by decreasing the drive to the ($i$+1)-st unit, with the side effect of increases in the active durations of the other uints.
Hence, a key difference of this model from the heteroclinic cycling case is the lack of a simple mechanism for independent phase modulation.
A perturbation can be implemented here such that one cell's active duration exhibits the predominant effect, but the timing of the other cells is still impacted.
This contrast in phase modulation properties is likely due to the lack of saddle points, which make an important contribution to the timing of trajectories in the heteroclinic cycle model.

In the relaxation oscillator cases, phase duration changes manifest through fast threshold modulation and associated timing effects.
Here, we can think of the units that are not directly perturbed as being unaffected (i.e., no change in flow or transition surfaces) except in the timing at which their transitions occur.  Consistent with past work \citep{daun2009control,rubin2009multiple}, however, the specific transition mechanisms do matter for what outcomes result from perturbations.
In the piecewise continuous heteroclinic case with state-dependent switching boundary surfaces (i.e., surfaces defined in terms of the dependent variables), the effects could arise both through changes in switching manifolds and through direct effects on the vector field.
To distinguish these factors, we also consider:
\begin{itemize}
 \item The piecewise-continuous heteroclinic case with fixed boundaries.  Here, we obtain almost the same results (not shown) as in the switching boundary case. This indicates that the dominant contribution to the duration changes comes from the vector field, especially the effect of changes in the vector field on the  positions where the trajectory reaches the surfaces of vector field discontinuity, not from changes in these surfaces themselves.
 \item The continuous heteroclinic case with fixed boundaries. This model has the fewest possible factors affecting the timing. We find that it has qualitatively the same responses (not shown) to perturbation as the piecewise-continuous cases.  Thus, the impact of a perturbation does not arise predominantly from the jumps in the vector field, but rather from the impact of the perturbed vector field on the trajectory position, irrespective of the vector field continuity.
 \end{itemize}
Unlike the relaxation oscillator cases, in the heteroclinic cycle and threshold-linear models, we cannot decompose the flow into lower-dimensional systems for separate units, and impacts on the flow occur throughout the path of the trajectory.  Interestingly, it is not obvious (at least to us) why these combinations of factors result in the diverse responses to perturbations that arise across the cases that we have considered.


A previous study performed a thorough analysis comparing effects of perturbations to single units within a fast-slow half-center oscillator model with either escape transitions, release transitions, or transitions via an intermediate ``adaptation'' mechanism combining elements of both escape and release \citep{daun2009control}. Interestingly, that study found that escape transitions restricted the impact of such perturbations to the silent duration of the perturbed unit or equivalently the active duration of the unperturbed unit, while release transitions led to compensatory effects and produced only weak changes in durations, and in the adaptation regime, one-sided perturbations  affected durations of both units.
In other words, independent \emph{silent} phase modulation was possible with escape transitions but not release transitions.
In fact, our results are consistent with these earlier findings.
In this study, in the escape regime, we do have independent modulation of silent phase durations.
Specifically, as shown in Table~\ref{tab:SE_duration_difference}, the  duration change in the silent phase 1 is given by $\Delta t_2+\Delta t_3$, which is significantly negative when a positive perturbation on $d_1$ is imposed, while the duration change in the silent phase 2, $\Delta t_1+\Delta t_3$, and the change in the silent phase 3, $\Delta t_1+\Delta t_2$, are negligible.
Moreover, in the intrinsic release regime, the larger duration change that we found is linked to the slowing down near the right knee of the voltage nullcline, such that the difference between the two papers in the release case comes not from the number of cells but rather from the position of the slow nullcline relative to the right knee.

\begin{table}
\centering
\caption{\label{tab:summary_results} Effects of increasing the drive to unit 1 on the phase durations $t_i$ and total period $T_0$ in each mechanism. Columns 1-3 and 4 list the dimensionless change $\frac{\Delta t_i}{t_i}\frac{d_1}{\Delta d_1}$ for each active phase duration, and $\frac{\Delta T_0}{T_0}\frac{d_1}{\Delta d_1}$ for total period.
Matrix $C$ for each case is circulant, with first row given by the elements of columns 1, 2, 3 of the table.
Column 4 gives the 2-norm of the inverse of $C$; column 5 gives the condition number of $C^{-1}$.}
\begin{tabular}{|c|c|c|c|c|c|c|}
\hline
&Unit 1&Unit 2&Unit 3&Total period&$\|C^{-1}\|$&$\text{cond}(C^{-1})$\\
\hline
Intrinsic release&+0.076&+0.000 &-0.000&+0.025&13.24&1.02\\
\hline
Synaptic release&+0.024&-0.000&+0.000&+0.008&42.00&1.02\\
\hline
Synaptic escape&+1.962&-1.920&-2.388&-0.782&0.43&1.76\\
\hline
Heteroclinic cycling&-0.048&-0.007&-0.316&-0.124&3.44&1.28\\
\hline
Threshold linear&+1.948&+1.708&-3.443&+0.071&4.69&24.77\\
\hline
\end{tabular}
\end{table}

Applying similar arguments, we can consider how well each model analyzed in this study achieves independent active phase modulation as could be important for downstream effectors driven by excitation from CPG units, independent silent phase modulation as could be important for downstream effects dependent on pauses in CPG unit activity, and either invariance or sensitivity of period to perturbations.
The information in Table \ref{tab:summary_results} allows us to discuss these properties heuristically and empirically, where for the latter, we define a circulant matrix $C$ for each case with first row given by the elements of columns 1, 2, 3 of the table.
We see that the most pronounced independent active phase modulation comes from the intrinsic release model, in which an input perturbation to cell 1 induces an increase in the cell 1 active phase duration that is strong relative to the negligible impacts on the other two cells' active phases.
As discussed above, the synaptic escape model achieves independent silent phase modulation.
The threshold-linear network also exhibits this property to some extent, because changes in input to unit 1 will produce much stronger impacts on the silent phase duration of unit 3 (comprising the active phase durations of units 1 and 2) than on those of the other units.
Finally, the ranking of the models in order of sensitivity of period to input modulation goes as follows:  synaptic escape, heteroclinic cycle, threshold-linear, intrinsic release, synaptic release.  This ranking aligns with $\|C^{-1}\|$ in Table \ref{tab:summary_results}.
Indeed, if a system wants to induce a change $\rho$ in its three active phase durations, then that corresponds to finding a vector $\mu$ of input strengths such that $\rho = C\mu$.
The size of the needed input is therefore bounded as $\|\mu\| \leq \|C^{-1}\|\|\rho\|$.
The threshold-linear model is interesting in that it allows for significant phase perturbations with relatively small changes in overall period, as evidenced by its large condition number (see $\text{cond}(C^{-1})$ in Table \ref{tab:summary_results}).
On the other hand, since all of the rhythm features in the synaptic release model show little response to input changes, that framework seems to be effective for robustness but not amenable to simple control; more generally, the relatively weak responses of the release models to input perturbations is consistent with past observations in HCOs \citep{daun2009control}.

On the other hand, a new result from our work that distinguishes the triphasic case from the HCO case
is the instability to perturbations for the triphasic rhythm with intrinsic escape transitions.  This finding does not rule out intrinsic escape as a transition mechanism for biological CPGs, which may feature asymmetries or differences in transition mechanisms for different phases as discussed below, but it does highlight the importance of checking robustness of rhythmic behavior in models, of paying attention to transition mechanisms in model design, and of taking care in extending small circuit models to larger ones.

A paradigmatic model for the brain stem neuronal network underlying respiratory rhythm generation includes three populations coupled with mutual inhibition, along with a fourth excitatory population. An investigation of that model based on fast-slow decomposition and bifurcation analysis revealed that the expiration-to-inspiration transition occurs via escape while the inspiration-to-expiration transition results from release.  Simulations yielded the predictions that varying a tonic control drive to the escaping inspiratory neurons would shorten the expiratory phase, while varying the drive to the released expiratory neurons would have relatively little effect on phase durations \citep{rubin2009multiple}. These results are consistent with the two-cell, half-center oscillator analysis \citep{daun2009control}, but it is important to note that the active phases of two of the three inhibitory respiratory populations overlap, so it is not surprising that this system acts more like a two-cell network than the type of three-cell network that we studied.  Although the results from the respiratory model, which combines escape and release, agreed qualitatively with those obtained from analysis of pure escape and pure release scenarios, it is not obvious that a similarly straightforward generalization of our analysis would yield correct predictions for three-cell networks that combine escape and release.

In contrast to these studies of effects of constant input levels on phase durations, Zhang and Lewis investigated differences in the infinitesimal phase response curves (iPRC) for HCO systems \citep{zhang2013phase}.
They found that the timing sensitivity of release systems to transient pertubations was greatest when applied to the active cell, near the end of the active phase.
Similarly, they found that escape systems showed greatest sensitivity to perturbations applied towards the end of an escaping cell's silent phase.
Here, we find behavior that is consistent with these observations.
In Figure~\ref{fig:lTRC_IR_SR}, for example, under the release mechanisms, the integrand of the local timing response curve integral was largest (greatest sensitivity) towards the end of the active phase of the cell (cell 1) subject to the perturbation (changing $d_1$).
In contrast, in Figure~\ref{fig:lTRC_SE}, under the escape mechanism, the largest values of the lTRC integrand occurred towards the end of the inactive phase of the cell coming one step earlier in the activation sequence than the cell subject to the parametric perturbation.

Extreme sensitivity of timing for limit cycle trajectories that pass close to a saddle or are near a heteroclinic or homoclinic bifurcation point has been noted previously.
For example, \cite{shaw2012phase} pointed out phase-dependent divergence of  infinitesimal phase response curves for a family of limit cycles that approach a heteroclinic cycle as a parameter is varied.
In another timing-related analysis, Izhikevich derived the phase equations for a coupled pair of relaxation oscillators and showed that in the singular limit, the effective coupling function exhibits discontinuous jumps at the points where fast phase transitions start and stop; these jumps perturb into large slopes when the singular perturbation parameter is small but nonzero \citep{izhikevich2000}.



This work offers a stepping stone towards a broader theory of control of limit cycle systems.
Control theory is most fully developed for linear systems \citep{brockett2015finite,stanhope2014identifiability,bechhoefer2021control}.
For a linear system such as $\dot{\mbx}=A\mbx+B\mbu$, with state vector $\mbx\in\R^n$, control input $\mbu\in\R^k$, and observable $\mby=C\mbx\in\R^m$,
the notions of controllability and observability are \emph{global}: one constructs the ``controllable" space $\mathcal{C}\subset\R^n$ of all points to which the system can be driven in finite time from initial $\mbx(0)=0$ via a suitable control, and the ``observable" space $\mathcal{O}\subset\R^n$ such that trajectories from different initial conditions within $\mathcal{O}$ may be distinguished by observing the trajectory $\mby(t)$.
For \emph{nonlinear} systems these notions are necessarily more limited, and are restricted to specific subclasses of systems \citep{isidori1985nonlinear,whalen2015observability}.
Here we consider nonlinear dynamical systems supporting stable limit cycle trajectories, as typically arise in central pattern generator models.
Previous work on control of limit cycle systems has focused on controlling global properties of limit cycles, for instance the amplitude or period of an ongoing oscillation, or (de)synchronization of a population of coupled oscillators \citep{monga2019phase,monga2019optimal,wilson2022recent}.
Here we focus on the control of specific subcomponents of a limit cycle trajectory, in motor control systems comprising three distinct phases of movement.
Classic analysis for understanding timing of control systems under a weak perturbation is well established by using the infinitesimal phase response curve (iPRC).
The iPRC can only capture the \emph{global} change in the timing induced by the perturbation, instead of the \emph{local} timing change in each subcomponent of the limit cycle.
To circumvent this limitation,  \cite{wang2021shape} developed the local timing response curve (lTRC), which allows one to compute local timing changes due to nonuniform timing sensitivities of the control system along the limit cycle.
In this work we have conducted analysis on the duration sensitivity in each phase of limit cycle systems by applying the lTRC, which yields excellent agreement with both direct numerial simulation and other mathematical analysis.
Our work provides detailed examples of how to analyze the responses to perturbations arising in multi-phase oscillations of naturally occurring biological motor control systems, by applying tools that allow linearized/variational analysis of the parametric variations.


From a biological point of view, a general important point evident in this and past work is that we cannot assume that effects of perturbation of one unit in a coupled circuit will be localized to that unit or even dominant in that unit.  Moreover, such effects may not materialize instantly; instead, we may need to observe one or more full cycles to see the persistent effects.
In addition, the lessons we extract from this work can provide insights into the selection of a CPG model for robotics or for representing biological systems.
Specifically, the divergent properties of the heteroclinic cycle CPG model and the relaxation oscillator CPG models can guide the selection of a suitable modeling framework to capture experimental observations or design aims for triphasic rhythmic systems, while similar analysis can be applied to other models producing other forms of multi-phase rhythms.

Note that circuits of interest for biological applications would typically not have the same degree of symmetry as those studied here.
For example, in the respiratory network, in addition to the possibility that both escape and release transitions arise during each respiratory cycle, it is unlikely that the effective strengths of all synaptic connections within the network are equal; indeed, these differ in models that meet experimental benchmarks \citep{golowasch1999network,latorre2002characterization,rubin2009multiple,rubin2011,park2022}. 
It remains an interesting open question to apply the methods developed here to more realistic circuits.
The work presented here serves as a template, although it remains to be seen to what extent the lessons learned here carry over to systems with heterogeneous switching mechanisms.
Similarly, the approaches in this work can be applied to analogous circuits with more than three components, for the study of solutions in which units' active phases do not overlap, albeit with additional bookkeeping.  What matters most for the analysis is the number of units influencing other units' dynamics and the location of switching surfaces at any one time. The relaxation oscillator framework with a steep or Heaviside-like synaptic coupling function $S_{\infty}(v)$ is particular conducive to the inclusion of more units (e.g., \cite{rubin2009multiple,molkov2015,webster2020control}). The precise results on tuning sensitivity associated with specific transition mechanisms may, however, depend on the number of units in the circuit, or possibly the parity of the circuit size; more units would also allow for the additional complication of diverse  coupling architectures.

In this paper we have considered deterministic CPG models.
In biological control circuits, stochastic fluctuations can also play an important role \citep{carroll2013cycle,yu2021dynamical,rubin2022irregular}.
For example, stochastic heteroclinic channels have been studied intensively \citep{stone1990random,armbruster2003noisy,bakhtin2011noisy}.
Horchler and colleagues used the amplitude of an additive noise signal as a control parameter to dynamically regulate the mean period of a three-population May-Leonard model \citep{horchler2015designing}.
It would be interesting to investigate how changing the amplitude of noise impinging on individual units affects the (mean) activation durations of all three units, in analogy to the single-input perturbations studied here, but this question goes beyond the scope of the current paper.
In his doctoral thesis, Shaw observed that one could manipulate the mean period of a stochastic 2D heteroclinic-channel based oscillator by changing the underlying vector field, by varying the additive noise level, or through a combination of both \citep{shaw2014dynamical}.
Results in \cite{Barendregt2022submitted} put this analysis on a rigorous footing and extend it to several variants of  the May-Leonard three-population model with discrete (non-Gaussian, non-additive) noise arising from demographic stochasticity.

Finally, we have presented three dynamical architectures that produce triphasic rhythms but our examples are certainly not exhaustive.
For instance, the repressilator is a genetic regulatory network consisting of three transcriptional repressor genes \citep{elowitz2000synthetic}.
The three units are connected in a negative feedback loop, allowing sustained limit-cycle oscillations.
Another example is the stomatogastric ganglion of decapod crustaceans.
Its constituent neurons encompass two central pattern generators, one for the gastric mill rhythm and one for the pyloric rhythm.
For the pyloric CPG, three nerves, containing all of the pyloric units, generate a triphasic rhythm.
This pattern controls striated muscles that dilate and constrict the pyloric region of the stomach, which moves the food caudally through filter combs inside the pylorus \citep{weimann1991neurons,selverston1998basic}.
The approaches we use here can be applied to these triphasic systems as well.

\subsection*{Acknowledgments}
The authors thank Yangyang Wang for helpful discussion about local timing response curves.
This work was supported in part by NSF grant DMS-1951095 (JER), by NIH grant 1 RF1 NS118606-01 (PJT) and by NSF grant DMS-2052109 (PJT).
PJT acknowledges research support from Oberlin College.

\appendix

\section{Model details}
\label{App:model_details}

In the relaxation oscillator system \eqref{eq:relaxation_equations}, the functions $x_\infty(v)$ for $x=h,m_p$ are given by
$$x_\infty(v)=\frac{1}{1+\exp{\left((v-\theta_x)/\sigma_x\right)}}.$$
Table~\ref{tab:parameter_values_ir} lists the parameter values for the intrinsic release transition mechanism, which produces Figure~\ref{fig:IR_solution}.
Based on Table~\ref{tab:parameter_values_ir}, change $\theta_\text{I}$ to -25 mV to generate the synaptic release mechanism (Figure~\ref{fig:SR_solution}); change $\theta_\text{I}$ to -62 mV and $\sigma_h$ to 5 to generate the synaptic escape mechanism (Figure~\ref{fig:SE_solution}).
For the intrinsic escape mechanism shown in Figure~\ref{fig:IE_solution_symmetric} and Figure~\ref{fig:IE_solution_asymmetric}, make the following changes:
\begin{align*}
    \theta_\text{I}=-36,\quad\theta_h=-39,\quad\sigma_h=9,\quad g_\text{I}=0.24,\quad g_\text{E}=0.14.
\end{align*}
The simulation codes are available
at https://github.com/zhuojunyu-appliedmath/Triphasic-control.
Instructions for reproducing each figure and table in the
paper are provided.

\begin{table}
\centering
\caption{\label{tab:parameter_values_ir} Parameter values for the relaxation oscillator model \eqref{eq:relaxation_equations} with the intrinsic-release transition mechanism (Figure~\ref{fig:IR_solution}).}
\begin{tabular}{|l|l|ll||l|l|ll||l|l|l|}
\hline
Parameter&Value&Unit&&Parameter&Value&Unit&&Parameter&Value&Unit\\
\hline
$C$&0.21&$\mu$F/$\text{cm}^2$&&$\theta_\text{I}$&-43&mV&&$\theta_h$&-40&mV\\
$\epsilon$&0.01&None&&$\sigma_\text{I}$&-0.01&None&&$\theta_\text{mp}$&-37&mV\\
$V_\text{Na}$&50&mV&&$g_\text{NaP}$&6.8&$\mu$S/$\text{cm}^2$&&$\sigma_h$&6&None\\
$V_\text{L}$&-65&mV&&$g_\text{L}$&3&$\mu$S/$\text{cm}^2$&&$\sigma_\text{mp}$&-6&None\\
$V_\text{I}$&-80&mV&&$g_\text{I}$&0.4&$\mu$S/$\text{cm}^2$&&$b_{ij}$&1&None\\
$V_\text{E}$&0&mV&&$g_\text{E}$&0.1&$\mu$S/$\text{cm}^2$&&$d_{i}$&as in text&None\\
\hline
\end{tabular}
\end{table}

\section{Local timing response curve (lTRC)}
\label{App:lTRC}

\subsection{Introduction to lTRCs}
\label{appssec:Details of lTRCs}

Consider a parametrized continuous-time dynamical system defined on a domain $\Omega\subset\mathbb{R}^n$,
\begin{equation}
    \label{eq:vector_field}
    \frac{dx}{dt}=F_\mu(x),
\end{equation}
where $x\in\Omega$, $\mu\in\R$ and $F_\mu(\cdot)$ is either smooth or piecewise smooth on $\Omega$.
We are interested in the case where $F_\mu(\cdot)$ admits a family of hyperbolically stable limit cycles
for some range of $\mu$ including $\mu=0$, which we consider the ``unperturbed" limit cycle $\gamma(t)$.
We assume the basins of attraction are contained in $\Omega$ and that the domain is partitioned into two or more subdomains, $\Omega=R^{\text{\uppercase\expandafter{\romannumeral1}}}\cup R^{\text{\uppercase\expandafter{\romannumeral2}}}\cup\cdots$
with successive domain boundaries transverse to the flow of the unperturbed limit cycle.
Moreover, we suppose that  $F_\mu(\cdot)$ is smooth (in $x$) within each domain.
(See \cite{park2018infinitesimal} for a fully detailed construction of such a system of domains.)
In the following we will write $F(x)$ for $F_0(x)$.
The local timing response curve (lTRC) introduced by \cite{wang2021shape} is defined to measure the timing sensitivity within each region in response to a static perturbation $\mu\not=0$.

For $x\in R^i,\,i=\text{\uppercase\expandafter{\romannumeral1},\uppercase\expandafter{\romannumeral2}},\cdots$, let $\Gamma^i(x)$ be the time remaining until the trajectory beginning at $x$  exits  the region.
At any point $x\in R^i$, by construction, $\Gamma^i$ satisfies
\begin{equation}
\label{eq:remaining_time}
\frac{d\Gamma^i(x)}{dt}=-1,\quad x\in R^i,
\end{equation}
along $\gamma(t)$. Thus, by the chain rule
\begin{equation}
\label{eq:F_dot_eta}
F(x)\cdot \nabla\Gamma^i(x(t))=-1,\quad x\in R^i.
\end{equation}
The associated lTRC for region $R^i$ is defined to be the gradient of $\Gamma^i$ evaluated along the limit cycle:
$$\eta^i(t)=\nabla\Gamma^i(\gamma(t)).$$
For convenience, we drop the superscript $i$ but note that every region is characterized by its own lTRC.
Differentiating both sides of \eqref{eq:F_dot_eta}, within any particular subdomain, with respect to $t$ gives
\begin{equation}
\label{eq:lTRC_ode}
    \frac{d\eta}{dt}=-DF(\gamma(t))^T\eta,
\end{equation}
where $DF$ denotes the Jacobian of $F$.
Let $x^\text{out}$ denote the exit point from region $R^i$ and $n^{\text{out}}$ denote a normal vector of the exit boundary of $R^i$ at $x^{\text{out}}$.
Then, following \eqref{eq:F_dot_eta}, the lTRC $\eta$ satisfies the boundary (normalization) condition
\begin{align}
\label{eq:lTRC_bd_condition}
    F(x^\text{out})\cdot \eta(x^\text{out})=-1\quad\Longrightarrow\quad\eta(x^\text{out})=-\frac{n^\text{out}}{(n^\text{out})^TF(x^\text{out})},
\end{align}
The adjoint equation \eqref{eq:lTRC_ode} together with the boundary (normalization) condition \eqref{eq:lTRC_bd_condition} defines the local timing response curve in each subdomain.

\subsection{Relationship between iPRC and lTRC}
\label{appssec:Releationship between iPRC and lTRC}

The formulation of the lTRC given by \eqref{eq:lTRC_ode} and \eqref{eq:lTRC_bd_condition} is similar to the formulation of the infinitesimal phase response curve (iPRC).
The iPRC, denoted by $z(t)$, satisfies the same adjoint equation
\begin{equation*}
\label{eq:iPRC_ode}
    \frac{dz}{dt}=-DF(\gamma(t))^T z,
\end{equation*}
with a different normalization condition
\begin{equation*}
\label{eq:iPRC_nor_condition}
    F(\gamma(t))\cdot z(t)=1.
\end{equation*}

\begin{figure}
\includegraphics[width=14.5cm]{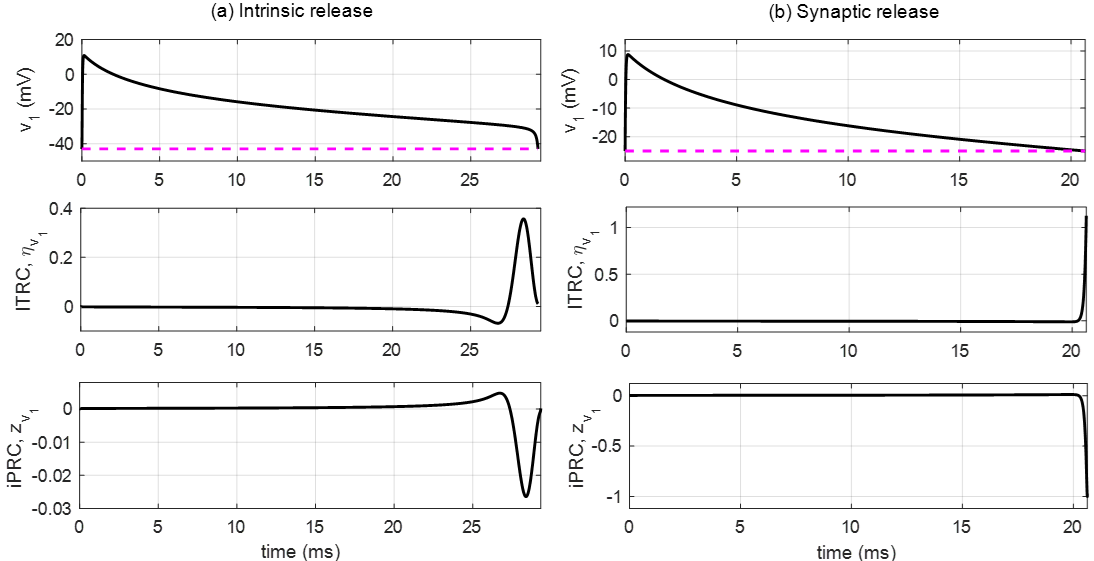}
\caption{\label{fig:iPRC_IR_SR} Comparison of the lTRC and iPRC during the active phase of cell 1 when $d_1$ is perturbed, with transitions by
\textbf{(a)} intrinsic release; \textbf{(b)} synaptic release. Top panels: time series of $v_1$; middle panels: lTRC $\eta$ in the $v_1$ direction; bottom panels: iPRC $z$ in the $v_1$ direction.
The lTRC and iPRC are symmetric in shape about the abscissa ($t$-axis).}
\end{figure}

Figure~\ref{fig:iPRC_IR_SR} shows the lTRC and iPRC during the first active phase for the relaxation-oscillator system with each of the two release mechanisms.
(Due to symmetry, the curves for phase 2 (or phase 3) are the same if the same perturbation is applied to $v_2$ (or $v_3$).)
As expected, the lTRC and iPRC have an identical geometric shape and are mutually symmetric about the time axis.

\subsection{Derivation of generalized $T_1$ formula \eqref{eq:T1_shifting_boundary}}
\label{appssec:Derivation of generalized T_1 formula}

This section presents the derivation of equation \eqref{eq:T1_shifting_boundary}, which is a generalization of \eqref{eq:T1_fixed_boundary} from the fixed-boundary case to the shifting-boundary case.
It follows the same generally exposition given in \cite{wang2021shape} but with some modifications.
Like the lTRC, the linear time shift $T_1^i$ is considered separately in each subdomain $R^i$, but the superscript $i$ is dropped for the time being.

\begin{figure}
\centering
\includegraphics[width=8.3cm]{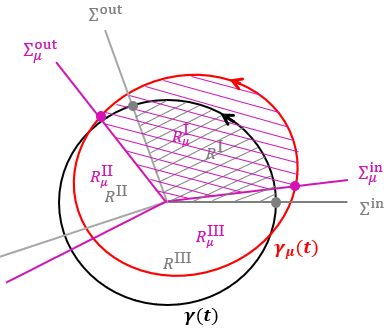}
\caption{\label{fig:shifting_boundaries} A schematic illustration of the shifting entry and exit surfaces for a system with three regions.
The unperturbed trajectory $\gamma(t)$ (black circle) enters region $R^{\text{\uppercase\expandafter{\romannumeral1}}}$ (grey shaded) through boundary $\Sigma^\text{in}$ and exits $R^{\text{\uppercase\expandafter{\romannumeral1}}}$ through boundary $\Sigma^\text{out}$;  the perturbed trajectory $\gamma_{\mu}(t)$ (red circle) enters region $R_\mu^{\text{\uppercase\expandafter{\romannumeral1}}}$ (magenta shaded) through boundary $\Sigma_{\mu}^\text{in}$ and exits $R_\mu^{\text{\uppercase\expandafter{\romannumeral1}}}$ through boundary $\Sigma_{\mu}^\text{out}$.
The grey dot on $\Sigma^\text{in}$ (resp., $\Sigma^\text{out}$) represents the entry position $x^\text{in}$ (resp., exit position $x^\text{out}$) of the unperturbed trajectory into (resp., from) region $R^{\text{\uppercase\expandafter{\romannumeral1}}}$.
The magenta dot on $\Sigma_\mu^\text{in}$ (resp., $\Sigma_\mu^\text{out}$) represents the entry position $x_\mu^\text{in}$ (resp., exit position $x_\mu^\text{out}$) of the perturbed trajectory into (resp., from) region $R_\mu^{\text{\uppercase\expandafter{\romannumeral1}}}$.}
\end{figure}

Suppose limit cycle $\gamma_{\mu}$ is a solution to the perturbed system
\begin{equation}
\label{eq:vector_field_perturbed}
    \frac{dx}{d\tau}=F_{\mu}(x),
\end{equation}
where $\tau=\tau(\mu,t)$ is the perturbed time coordinate satisfying
$$\tau(0,t)=t,\quad\tau(\mu,t+T_0)=\tau(\mu,t)+T_{\mu},$$
where $T_0$ and $T_{\mu}$ represent the durations spent by  the unperturbed trajectory and the perturbed trajectory, respectively, in region $R^i$.
For simplicity, consider the linear scaling
\begin{equation}
\label{eq:rescaled_time}
    \frac{d\tau}{dt}=\frac{1}{\nu_{\mu}}=\frac{T_{\mu}}{T_0}.
\end{equation}
Suppose the unperturbed trajectory $\gamma(t)$ enters $R^i$ through boundary $\Sigma^\text{in}$ at $x^\text{in}$ when $t=t^\text{in}$, and it exits the region through $\Sigma^\text{out}$ at $x^\text{out}$ when $t=t^\text{out}$; we use similar notation for the perturbed trajectory $\gamma_{\mu}(\tau)$.
Assume the boundaries are not fixed, but perturb to $\Sigma_{\mu}^\text{in}$ and $\Sigma_{\mu}^\text{out}$ under the static perturbation $\mu$, as illustrated in Figure~\ref{fig:shifting_boundaries}.
Without loss of generality, let $t^\text{out}=0$, and thus, the time duration the limit cycles spend in $R^i$ is $$T_0=-t^\text{in},\qquad T_{\mu}=t_{\mu}^\text{out}-t_{\mu}^\text{in}.$$
Therefore, at entry and exit points to region $R^i$, the following expressions give the times that the unperturbed and perturbed limit cycles have left to spend in region $R^i$:
\begin{align*}
    &\Gamma(x^\text{in})=-t^\text{in}=T_0,\quad \Gamma(x^\text{out})=0,\\
    &\Gamma_{\mu}(x_{\mu}^\text{in})=t_{\mu}^\text{out}-t_{\mu}^\text{in}=T_{\mu},\quad \Gamma_{\mu}(x_{\mu}^\text{out})=0.
\end{align*}
As in \eqref{eq:remaining_time} and \eqref{eq:F_dot_eta}, for any $\mu,$
\begin{equation}
\label{eq:normalization}
    -1=\frac{d\Gamma_{\mu}(\gamma_{\mu}(\tau))}{d\tau}=F_\mu(\gamma_{\mu}(\tau))\cdot\eta_{\mu}(\gamma_{\mu}(\tau)).
\end{equation}
Integrating \eqref{eq:normalization} from $\tau=t_{\mu}^\text{out}$ to $t_{\mu}^\text{in}$ gives
\begin{equation}
\label{eq:T_epsilon}
    T_{\mu}=\int_{t_{\mu}^\text{out}}^{t_{\mu}^\text{in}}F_\mu(\gamma_{\mu}(\tau))\cdot\eta_{\mu}(\gamma_{\mu}(\tau))\,d\tau.
\end{equation}
On the other hand, we can expand $T_{\mu}$ as
\begin{equation}
\label{eq:T_epsilon_expansion}
    T_{\mu}=T_0+\mu T_1+O({\mu}^2),
\end{equation}
where $T_1$ is the first-order change in the duration the trajectory spends in region $R^i$ resulting from perturbation $\mu$.
Our goal is to obtain the expression for $T_1$ using Taylor expansions for all of the terms in \eqref{eq:T_epsilon}.

Suppose we can expand $F_{\mu}$, $\Gamma_{\mu}$, $\eta_{\mu}$ and $\gamma_{\mu}$ as follows:
\begin{equation}
\label{eq:Taylor}
\begin{split}
    F_{\mu}(x)&=F(x)+\mu F_1(x)+O(\mu^2),\\
    \Gamma_{\mu}(x)&=\Gamma(x)+\mu\Gamma_1(x)+O(\mu^2),\\
    \eta_{\mu}(x)&=\eta(x)+\mu\eta_1(x)+O(\mu^2),\\
    \gamma_{\mu}(\tau)&=\gamma(t)+\mu\gamma_1(t)+O(\mu^2).
\end{split}
\end{equation}
From \eqref{eq:rescaled_time} and the exit time expressions, the relation between $t$ and $\tau$ is
\begin{equation}
\label{eq:relation between t and tau}
    t=\nu_{\mu}(\tau-t_{\mu}^{\text{out}})=\frac{T_0}{T_{\mu}}(\tau-t_{\mu}^{\text{out}}).
\end{equation}
Accordingly, expand \eqref{eq:T_epsilon} to first order:
\begin{align*}
    T_{\mu}=&\int_{t_{\mu}^\text{out}}^{t_{\mu}^\text{in}}\left[F(\gamma(t(\tau)))+\mu DF(\gamma(t(\tau)))\cdot\gamma_1(t(\tau))+\mu F_1(\gamma(t(\tau)))\right]\cdot\\
    &\qquad\left[\eta(\gamma(t(\tau))+\mu D\eta(\gamma(t(\tau)))\cdot\gamma_1(t(\tau))+\mu\eta_1(\gamma(t(\tau)))\right]\,d\tau+O(\mu^2)\\
    =&\int_{t_{\mu}^\text{out}}^{t_{\mu}^\text{in}}\left\{ F(\gamma(t(\tau)))\cdot\eta(\gamma(t(\tau))+\mu\left[F(\gamma(t(\tau)))\cdot\eta_1(\gamma(t(\tau)))+F_1(\gamma(t(\tau)))\cdot\eta(\gamma(t(\tau))\right] \right.+\\
    &\left. \qquad\mu\left[F(\gamma(t(\tau)))\cdot D\eta(\gamma(t(\tau)))\cdot\gamma_1(t(\tau))+DF(\gamma(t(\tau)))\cdot\gamma_1(t(\tau))\cdot\eta(\gamma(t(\tau))\right]\right\} \,d\tau+O(\mu^2) \\
    \overset{\eqref{eq:relation between t and tau}}{=}&\frac{1}{\nu_{\mu}}\int_{0}^{-T_0}\left\{ F(\gamma(t))\cdot\eta(\gamma(t))+\mu\left[F(\gamma(t))\cdot\eta_1(\gamma(t))+F_1(\gamma(t))\cdot\eta(\gamma(t))\right]+ \right. \\
    &\left. \qquad\qquad\qquad\mu\left[F(\gamma(t)\cdot D\eta(\gamma(t))\cdot\gamma_1(t)+DF(\gamma(t))\cdot\gamma_1(t)\cdot\eta(\gamma(t))\right]\right\} \, dt +O(\mu^2).
\end{align*}
At $\mu=0$, \eqref{eq:T_epsilon} gives
$$T_0=\int_{0}^{-T_0}F(\gamma(t))\cdot\eta(\gamma(t))\,dt,$$
and \eqref{eq:rescaled_time} gives $T_0/\nu_{\mu}=T_{\mu}$.
Thus, the above equation becomes
\begin{scriptsize}
\begin{align*}
    0=&\int_{0}^{-T_0}\left\{ \left[F(\gamma(t))\cdot\eta_1(\gamma(t))+F_1(\gamma(t))\cdot\eta(\gamma(t))\right]+\left[F(\gamma(t))\cdot D\eta(\gamma(t))\cdot\gamma_1(t)+DF(\gamma(t))\cdot\gamma_1(t)\cdot\eta(\gamma(t))\right]\right\}\,dt+O(\mu)\\
    =&\int_{0}^{-T_0}\left\{ \left[F(\gamma(t))\cdot\eta_1(\gamma(t))+F_1(\gamma(t))\cdot\eta(\gamma(t))\right]+\left[F(\gamma(t))^T D\eta(\gamma(t))+\eta(\gamma(t))^T DF(\gamma(t))\right]\cdot\gamma_1(t)\right\} \,dt+O(\mu).
\end{align*}
\end{scriptsize}
Following \eqref{eq:normalization},
\begin{align*}
    &\qquad\quad F\cdot\eta=\sum_i\eta^iF^i=-1\\
    &\Longrightarrow\;\frac{\partial}{\partial x_j}\left(\sum_i\eta^iF^i\right)=\sum_i\frac{\partial\eta^i}{\partial x_j}F^i+\sum_i\eta^i\frac{\partial F^i}{\partial x_j}=0\\
    &\Longrightarrow\;F^T D\eta+\eta^T DF=0.
\end{align*}
Thus, we have
\begin{align}
\label{eq:two_terms}
    0=&\int_{0}^{-T_0}\left[F(\gamma(t))\cdot\eta_1(\gamma(t))+F_1(\gamma(t))\cdot\eta(\gamma(t))\right]\,dt+O(\mu).
\end{align}
Since
$$F(\gamma(t))=\frac{d\gamma(t)}{dt} \; \mbox{and} \; \quad\eta_1(x)=\frac{\partial\eta_{\mu}(x)}{\partial \mu}\bigg|_{\mu=0}=\frac{\partial\nabla\Gamma_{\mu}(x)}{\partial\mu}\bigg|_{\mu=0},$$
we have
\begin{align*}
    \int_{0}^{-T_0}F(\gamma(t))\cdot\eta_1(\gamma(t))\,dt&=\int_{t=0}^{-T_0}\frac{d\gamma(t)}{dt}\cdot\frac{\partial\nabla\Gamma_{\mu}(\gamma(t))}{\partial\mu}\bigg|_{\mu=0}\,dt\\
    &=\int_{0}^{-T_0}\frac{d\gamma(t)}{dt}\cdot\nabla\left(\frac{\partial\Gamma_{\mu}(\gamma(t))}{\partial\mu}\right)\bigg|_{\mu=0}\,dt\\
    &=\int_{0}^{-T_0}\frac{d}{dt}\left(\frac{\partial\Gamma_{\mu}}{\partial\mu}(\gamma(t))\right)\bigg|_{\mu=0}\,dt\\
    &=\left(\frac{\partial\Gamma_{\mu}}{\partial\mu}(x^\text{in})\right)\bigg|_{\mu=0}-\left(\frac{\partial\Gamma_{\mu}}{\partial\mu}(x^\text{out})\right)\bigg|_{\mu=0}\\
    &=\Gamma_1(x^\text{in})-\Gamma_1(x^\text{out}).
\end{align*}

Note that the key difference in this derivation relative to the previously derived  fixed-boundary case is that in the fixed-boundary case, $\frac{\partial\Gamma_{\mu}(x^\text{out})}{\partial\mu}\big|_{\mu=0}=0$.
From the above calculation, we can simplify \eqref{eq:two_terms} as
\begin{align}
    0&=\Gamma_1(x^\text{in})-\Gamma_1(x^\text{out})+\int_{0}^{-T_0}F_1(\gamma(t))\cdot\eta(\gamma(t))\,dt+O(\epsilon)\nonumber\\
    \Gamma_1(x^\text{in})&=\Gamma_1(x^\text{out})+\int_{-T_0}^{0}F_1(\gamma(t))\cdot\eta(\gamma(t))\,dt+O(\epsilon)\nonumber\\
    &=\Gamma_1(x^\text{out})+\int_{t_0^\text{in}}^{t_0^\text{out}}F_1(\gamma(t))\cdot\eta(\gamma(t))\,dt+O(\epsilon).
\label{eq:Gamma_1(x_0^in)}
\end{align}
Moreover, by the Taylor expansions given in \eqref{eq:Taylor},
\begin{align*}
    T_{\mu}&=\Gamma_{\mu}(x_{\mu}^\text{in})=\Gamma(x_{\mu}^\text{in})+\mu\Gamma_1(x_{\mu}^\text{in})+O(\mu^2)\\
    &=\Gamma(x^\text{in})+\mu\nabla\Gamma(x^\text{in})\cdot x_1^\text{in}+\mu\Gamma_1(x^\text{in})+O(\mu^2)\\
    &=\Gamma(x^\text{in})+\mu(\eta(x^\text{in})\cdot x_1^\text{in}+\Gamma_1(x^\text{in}))+O(\mu^2),
\end{align*}
where $x_1^\text{in}=\frac{\partial x_{\mu}^\text{in}}{\partial\mu}\big|_{\mu=0}.$
Equate the first order terms of $T_\mu$ with $T_1$ and substitute $\Gamma_1(x^\text{in})$ as given in \eqref{eq:Gamma_1(x_0^in)}:
\begin{align*}
    T_1&=\eta(x^\text{in})\cdot x_1^\text{in}+\Gamma_1(x^\text{in})\\
    &=\eta(x^\text{in})\cdot x_1^\text{in}+\Gamma_1(x^\text{out})+\int_{t^\text{in}}^{t^\text{out}}F_1(\gamma(t))\cdot\eta(\gamma(t))\,dt.
\end{align*}
Now only $\Gamma_1(x^\text{out})$ remains to be defined. Similarly, the expansion of $\Gamma_{\mu}(x_{\mu}^\text{out})$ gives
$$\Gamma_{\mu}(x_{\mu}^\text{out})=\Gamma(x^\text{out})+\mu(\eta(x^\text{out})\cdot x_1^\text{out}+\Gamma_1(x^\text{out}))+O(\mu^2).$$
Note that $\Gamma_{\mu}(x_{\mu}^\text{out})=\Gamma(x^\text{out})=0$, so
$$\Gamma_1(x^\text{out})=-\eta(x^\text{out})\cdot x_1^\text{out}+O(\mu).$$
Thus, we finally obtain the expression for the first-order change in the duration through region $R^i$:
\begin{equation*}
\begin{split}
    T_1&=\eta(x^\text{in})\cdot x_1^\text{in}-\eta(x^\text{out})\cdot x_1^\text{out}+\int_{t^\text{in}}^{t_0^\text{out}}F_1(\gamma(t))\cdot\eta(\gamma(t))\,dt\\
    &=\eta(x^\text{in})\cdot\frac{\partial x_{\mu}^\text{in}}{\partial\mu}\bigg|_{\mu=0}-\eta(x^\text{out})\cdot\frac{\partial x_{\mu}^\text{out}}{\partial\mu}\bigg|_{\mu=0}+\int_{t^\text{in}}^{t^\text{out}}\eta(\gamma(t))\cdot\frac{\partial F_{\mu}(\gamma(t))}{\partial\mu}\bigg|_{\mu=0}\,dt,
\end{split}
\end{equation*}
which is exactly equation \eqref{eq:T1_shifting_boundary}.

\section{Instability for intrinsic escape}
\label{app:IE_instability}

\begin{figure}
\centering
\includegraphics[width=14.5cm]{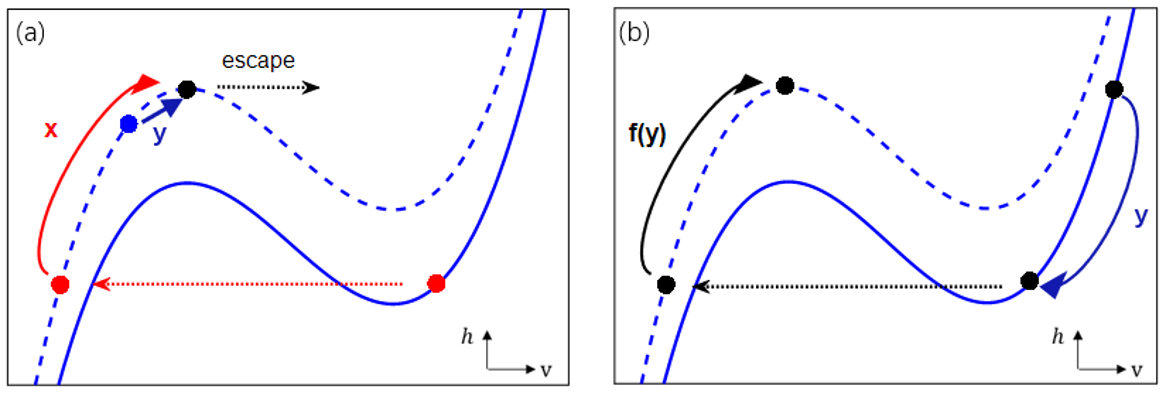}%
\caption{\label{fig:IE_sketch} A schematic of the cells in the intrinsic escape mechanism in the singular limit $\epsilon\rightarrow0$. The dashed blue line and solid blue line represent the inhibited $v$-nullcline and free $v$-nullcline, respectively. \textbf{(a)}: cell 1 (black dot) is about to escape; cell 3 (red dot) is at the jump-down position and ready to jump to the lower silent position; cell 2 (blue dot) is at the upper silent position. The passage time in the silent phase of cell 3 and cell 2 is denoted by $x$ and $y$, respectively. \textbf{(b)}: function $f$ that maps the active duration $y$ (of cell 1 here) to the silent duration $f(y)$.}
\end{figure}

This section gives an analysis of the instability of the equal-duration firing pattern for the intrinsic-escape transition mechanism.
We consider the case in the singular limit $\epsilon\rightarrow0$, as sketched in Figure~\ref{fig:IE_sketch}.
(Note that compared with Figure~\ref{fig:IE_solution_symmetric}, there is no double-inhibition notch.)

In this activity pattern, all cells jump up from the left knee of the inhibited nullcline.
Suppose cell 1 (black) is ready to escape and jump up. Let $x$ denote the passage time of cell 3 (red) from its jump-down position to its escape position (left knee of the inhibited $v_3$-nullcline).
Let $y$ denote the passage time of cell 2 (blue) from its initial position to the escape position.
Correspondingly, the time cell 1 spends in the active phase is equal to $y$.
Since all cells jump up from a specific position, we can
define the map $f(y)$ that translates the time $y$ that a cell spends in the active phase to the time that it spends in its next silent phase.

Each phase within a full evolution cycle starts with an analogous configuration, with a ``lower silent'' cell that just jumped down, an ``upper silent'' cell that will be next to escape, and an escaping cell.
Table~\ref{tab:IE_passage time}  gives the passage time for the two cells in the silent phase as time passes; the left column denotes the total time spent in the silent phase while the right column denotes the time spent in the silent phase from the jump up of a cell's predecessor to the jump up of that cell.

\begin{table}
\centering
\caption{\label{tab:IE_passage time} Intrinsic escape when $\epsilon\rightarrow0$: time of passage for lower and upper silent cells.}
\begin{tabular}{|c|c|c|}
\hline
&\makecell[c]{Passage time of\\lower silent cell}&\makecell[c]{Passage time of\\upper silent cell}\\
\hline
cycle 1&$x$&$y$\\
\hline
cycle 2&$f(y)$&$x-y$\\
\hline
cycle 3&$f(x-y)$&$f(y)-x+y$\\
\hline
cycle 4&$f(f(y)-x+y)$&$f(x-y)-(f(y)-x+y)$\\
\hline
$\vdots$&$\vdots$&$\vdots$\\
\hline
\end{tabular}
\end{table}

We are interested in a fixed point of the map defined from cycle 1 to cycle 4:
\begin{equation}
\label{eq:passage_time_map}
    (x,y)\longmapsto(f(f(y)-x+y),\,f(x-y)-(f(y)-x+y)),
\end{equation}
with $x=2y$ and $f(y)=x$, so that all the three phases have equal durations.
We observe that $f'(y)\in(0,1)$, because based on the slowing rate of flow as a trajectory approaches the right knee, a small increase in $y$ translates to a very small decrease in the position of the active cell at jump-down and hence a very small increase in $f(y)$.
With this observation, we check the stability of the equal-duration oscillation using the Jacobian of the map \eqref{eq:passage_time_map},
$$J=\left(\begin{matrix}-f'(f(y)-y)&(f'(y)+1)f'(f(y)-y)\\
f'(y)+1&-2f'(y)-1\end{matrix}\right).$$
It is easy to see that $\text{tr}(J)<0$ and $\text{det}(J)<0$.
Hence, the desired equal-duration solution, if it exists, is unstable.

\section{Analysis of duration changes for model \eqref{eq:Aplysia_equations}}
\label{app:analytical_duration_heteroclinic_cycling}

In this section, we provide an analytical formula to approximate the first-order duration change in each phase of the heteroclinic cycling model \eqref{eq:Aplysia_equations}.
In phase 1, pools $x$ and $y$ satisfy
\begin{align*}
    \frac{dx}{dt}&=1-x-(y+a_1)\rho,\quad x(0)=x_0,\\
    \frac{dy}{dt}&=y+a_2,\quad y(0)=y_0,
\end{align*}
and we obtain
\begin{equation}
\label{eq:xy_phase1_app}
\begin{split}
    y(t)&=y_0e^t+a_2(e^t-1),\\
    x(t)&=e^{-t}\left[x_0-1+(y_0-a_2)\rho/2\right]-e^t(y_0+a_2)\rho/2+1+a_2\rho+a_1(e^{-t}-1)\rho.
\end{split}
\end{equation}
With perturbation $a_1\rightarrow a_1+\mu$, denote the perturbed $(x,y)$-solution and the unperturbed $(x,y)$-solution to be $(x_p(t),y_p(t))$ and $(x_u(t),y_u(t))$, respectively.
 From \eqref{eq:xy_phase1_app},
\begin{align}
y_p(t)&=y_u(t)+e^t(y_p^0-y_u^0),\label{eq:yp1_yu1_relation_app}\\
x_p(t)&=x_u(t)+e^{-t}(x^0_p-x_u^0)+(e^{-t}-e^t)(y_p^0-y_u^0)\rho/2+\mu(e^{-t}-1)\rho\label{eq:xp1_xu1_relation_app},
\end{align}
where $(x_p^0,y_p^0)$ and $(x_u^0,y_u^0)$ represent the entry points for the perturbed case and unperturbed case, respectively.
Let $t_1$ denote the unperturbed exit time and $t_1^*=t_1+\Delta t_1$ the perturbed exit time; from the above observations, $\Delta t_1<0$.
On the exit surface,
\begin{align}
    x_p(t_1+\Delta t_1)&=y(t_1+\Delta t_1)+\frac{a_1+\mu+a_2}{2},\label{eq:exit1_new_app}\\
    x_u(t_1)&=y(t_1)+\frac{a_1+a_2}{2}\label{eq:exit1_old_app}.
\end{align}
Substituting \eqref{eq:yp1_yu1_relation_app} and \eqref{eq:xp1_xu1_relation_app} into \eqref{eq:exit1_new_app} yields
\begin{align*}
    x_u(t_1+\Delta t_1)+e^{-(t_1+\Delta t_1)}(x^0_p-x_u^0)+(e^{-(t_1+\Delta t_1)}-e^{t_1+\Delta t_1})(y_p^0-y_u^0)\rho/2+\mu(e^{-(t_1+\Delta t_1)}-1)\rho\\
    =y_u(t_1+\Delta t_1)+e^{t_1+\Delta t_1}(y_p^0-y_u^0)+\frac{\mu}{2}+\frac{a_1+a_2}{2}.
\end{align*}
By Taylor expansion of $x_u(t_1+\Delta t_1)$, $y(t_1+\Delta t_1)$, $e^{t_1+\Delta t_1}$ and $e^{-(t_1+\Delta t_1)}$ at $t=t_1$ and substitution of \eqref{eq:exit1_old_app}, we obtain
\begin{align}
\label{eq:xu'_y'_app}
    x_u'(t_1)\Delta t_1+e^{-t_1}(1-\Delta t_1)(x^0_p-x_u^0)+[e^{-t_1}(1-\Delta t_1)-e^{t_1}(1+\Delta t_1)](y_p^0-y_u^0)\rho/2+\nonumber\\
    \mu[e^{-t_1}(1-\Delta t_1)-1]\rho
    =y_u'(t_1)\Delta t_1+e^{t_1}(1+\Delta t_1)(y_p^0-y_u^0)+\frac{\mu}{2}+O(\Delta t_1^2).
\end{align}
Using the ODEs for the $x$ and $y$ variables and the fact that $a_1=a_2$ gives
\begin{align*}
    y_u'(t_1)&=y_u(t_1)+a_2=x_u(t_1)-\frac{a_1+a_2}{2}+a_2=x_u(t_1),\\
    x_u'(t_1)&=1-x_u(t_1)-(y_u(t_1)+a_1)\rho\\
    &=1-x_u(t_1)-\left(x_u(t_1)-\frac{a_1+a_2}{2}+a_1\right)\rho\\
    &=1-(1+\rho)x_u(t_1).
\end{align*}
Then, collecting the $\Delta t_1$ terms and the constant terms in \eqref{eq:xu'_y'_app} gives us a first-order estimate of $\Delta t_1$
\begin{align}
\label{eq:calculate_t1_app}
    \Delta t_1=\frac{\mu[1/2+(1-e^{-t_1})\rho]-e^{-t_1}(x^0_p-x_u^0)-(e^{-t_1}-e^{t_1})(y_p^0-y_u^0)\rho/2+e^{t_1}(y_p^0-y_u^0)}{1-(2+\rho)x_u(t_1)-\mu e^{-t_1}\rho-e^{-t_1}(x^0_p-x_u^0)-(e^{-t_1}+e^{t_1})(y_p^0-y_u^0)\rho/2-e^{t_1}(y_p^0-y_u^0)}+O(\Delta t_1^2)
\end{align}

In phase 3, the dynamics of pools $x$ and $z$ satisfies
\begin{align*}
    \frac{dx}{dt}&=x+a_1,\quad x(0)=x_0,\\ \frac{dz}{dt}&=1-z-(x+a_3)\rho,\quad z(0)=z_0,
\end{align*}
and we obtain
\begin{align*}
    x(t)&=x_0e^t+a_1(e^t-1),\\
    z(t)&=e^{-t}\left[z_0-1+(2a_3+x_0)\rho/2\right]-e^tx_0\rho/2+1-a_3\rho+a_1(2-e^t-e^{-t})\rho/2.
\end{align*}
Similar to phase 1,
\begin{align*}
    x_p(t)&=x_u(t)+e^t(x_p^0-x_u^0)+\mu(e^t-1),\\
    z_p(t)&=z_u(t)+e^{-t}(z_p^0-z_u^0)+(e^{-t}-e^t)(x_p^0-x_u^0)\rho/2+\mu(2-e^t-e^{-t})\rho/2,\\
    z_p(t_3+\Delta t_3)&=x_p(t_3+\Delta t_3)+\frac{a_1+\mu+a_3}{2},\\
    z_u(t_3)&=x_u(t_3)+\frac{a_1+a_3}{2}.
\end{align*}
Substituting the first two equations into the third equation and using Taylor expansion, with cancellation from the last equation, yields
\begin{align*}
    z_u'(t_3)\Delta t_3+e^{-t_3}(1-\Delta t_3)(z_p^0-z_u^0)+[e^{-t_3}(1-\Delta t_3)-e^{t_3}(1+\Delta t_3)](x_p^0-x_u^0)\rho/2\\
    +\mu[2-e^{t_3}(1+\Delta t_3)-e^{-t_3}(1-\Delta t_3)]\rho/2
    =x'_u(t_3)\Delta t_3+e^{t_3}(1+\Delta t_3)(x_p^0-x_u^0)+\\\Delta a_1[e^{t_3}(1+\Delta t_3)-1]+\frac{\mu}{2}+O(\Delta t_3^2).
\end{align*}
Note that
\begin{align*}
    x'_u(t_3)&=x_u(t_3)+a_1,\\
    z'_u(t_3)&=1-z_u(t_3)-(x_u(t_3)+a_3)\rho\\
    &=1-\left(x_u(t_3)+\frac{a_1+a_3}{2}\right)-(x_u(t_3)+a_3)\rho\\
    &=1-(1+\rho)x_u(t_3)-a_3\rho-\frac{a_1+a_3}{2}.
\end{align*}
Thus, we obtain
\begin{scriptsize}
\begin{align}
\label{eq:calculate_t3_app}
    \Delta t_3=\frac{\mu(1/2+\rho)+e^{-t_3}(z_p^0-z_u^0)+e^{-t_3}(x_p^0-x_u^0-\mu)\rho/2-e^{t_3}(x_p^0-x_u^0+\mu)(1+\rho/2)}{a_3\rho-1+(2+\rho)x_u(t_3)+e^{-t_3}(z_p^0-z_u^0)+e^{-t_3}(x_p^0-x_u^0-\mu)\rho/2+e^{t_3}(x_p^0-x_u^0+\mu)(1+\rho/2)+\frac{3a_1+a_3}{2}}+O(\Delta t_3^2).
\end{align}\end{scriptsize}
Both of the analytical approximation \eqref{eq:calculate_t1_app} and \eqref{eq:calculate_t3_app} have a great agreement with direct numerical calculation and lTRC calculation (see \S\ref{ssec:sensitivity_HC}).

\bibliography{bib_file}

\begin{thebibliography}{}

\bibitem[Afraimovich et~al., 2004]{afraimovich2004heteroclinic}
Afraimovich, V.~S., Rabinovich, M.~I., and Varona, P. (2004).
\newblock Heteroclinic contours in neural ensembles and the winnerless
  competition principle.
\newblock {\em International Journal of Bifurcation and Chaos},
  14(04):1195--1208.

\bibitem[Armbruster et~al., 2003]{armbruster2003noisy}
Armbruster, D., Stone, E., and Kirk, V. (2003).
\newblock Noisy heteroclinic networks.
\newblock {\em Chaos: An Interdisciplinary Journal of Nonlinear Science},
  13(1):71--79.

\bibitem[Bakhtin, 2011]{bakhtin2011noisy}
Bakhtin, Y. (2011).
\newblock Noisy heteroclinic networks.
\newblock {\em Probability theory and related fields}, 150(1):1--42.

\bibitem[Barendregt and Thomas, 2022]{Barendregt2022submitted}
Barendregt, N.~W. and Thomas, P.~J. (2022).
\newblock Heteroclinic cycling and extinction in may-leonard models with
  demographic stochasticity.
\newblock {\em submitted}.

\bibitem[Bechhoefer, 2021]{bechhoefer2021control}
Bechhoefer, J. (2021).
\newblock {\em Control Theory for Physicists}.
\newblock Cambridge University Press.

\bibitem[Bertram and Rubin, 2017]{bertram2017}
Bertram, R. and Rubin, J.~E. (2017).
\newblock Multi-timescale systems and fast-slow analysis.
\newblock {\em Mathematical Biosciences}, 287:105--121.

\bibitem[Brockett, 2015]{brockett2015finite}
Brockett, R.~W. (2015).
\newblock {\em Finite dimensional linear systems}.
\newblock Society for Industrial and Applied Mathematics.

\bibitem[Brown et~al., 2004]{brown2004phase}
Brown, E., Moehlis, J., and Holmes, P. (2004).
\newblock On the phase reduction and response dynamics of neural oscillator
  populations.
\newblock {\em Neural computation}, 16(4):673--715.

\bibitem[B{\"u}schges et~al., 2008]{buschges2008}
B{\"u}schges, A., Akay, T., Gabriel, J.~P., and Schmidt, J. (2008).
\newblock Organizing network action for locomotion: insights from studying
  insect walking.
\newblock {\em Brain research reviews}, 57(1):162--171.

\bibitem[Butera~Jr et~al., 1999a]{butera1999models1}
Butera~Jr, R.~J., Rinzel, J., and Smith, J.~C. (1999a).
\newblock Models of respiratory rhythm generation in the pre-botzinger complex.
  i. bursting pacemaker neurons.
\newblock {\em Journal of neurophysiology}, 82(1):382--397.

\bibitem[Butera~Jr et~al., 1999b]{butera1999models2}
Butera~Jr, R.~J., Rinzel, J., and Smith, J.~C. (1999b).
\newblock Models of respiratory rhythm generation in the pre-botzinger complex.
  ii. populations of coupled pacemaker neurons.
\newblock {\em Journal of neurophysiology}, 82(1):398--415.

\bibitem[Carroll and Ramirez, 2013]{carroll2013cycle}
Carroll, M.~S. and Ramirez, J.-M. (2013).
\newblock Cycle-by-cycle assembly of respiratory network activity is dynamic
  and stochastic.
\newblock {\em Journal of neurophysiology}, 109(2):296--305.

\bibitem[Daun and B{\"u}schges, 2011]{daun2011}
Daun, S. and B{\"u}schges, A. (2011).
\newblock From neuron to behavior: dynamic equation-based prediction of
  biological processes in motor control.
\newblock {\em Biological cybernetics}, 105(1):71--88.

\bibitem[Daun et~al., 2009]{daun2009control}
Daun, S., Rubin, J.~E., and Rybak, I.~A. (2009).
\newblock Control of oscillation periods and phase durations in half-center
  central pattern generators: a comparative mechanistic analysis.
\newblock {\em Journal of computational neuroscience}, 27(1):3--36.

\bibitem[Elowitz and Leibler, 2000]{elowitz2000synthetic}
Elowitz, M.~B. and Leibler, S. (2000).
\newblock A synthetic oscillatory network of transcriptional regulators.
\newblock {\em Nature}, 403(6767):335--338.

\bibitem[Ermentrout and Terman, 2010]{ermentrout2010}
Ermentrout, B. and Terman, D.~H. (2010).
\newblock {\em Mathematical foundations of neuroscience}, volume~35.
\newblock Springer.

\bibitem[Golowasch et~al., 1999]{golowasch1999network}
Golowasch, J., Casey, M., Abbott, L., and Marder, E. (1999).
\newblock Network stability from activity-dependent regulation of neuronal
  conductances.
\newblock {\em Neural computation}, 11(5):1079--1096.

\bibitem[Grodins et~al., 1954]{grodins1954respiratory}
Grodins, F.~S., Gray, J.~S., Schroeder, K.~R., Norins, A.~L., and Jones, R.~W.
  (1954).
\newblock Respiratory responses to {{CO2}} inhalation. {{A}} theoretical study
  of a nonlinear biological regulator.
\newblock {\em Journal of applied physiology}, 7(3):283--308.

\bibitem[Hao et~al., 2011]{hao2011}
Hao, Z.-Z., Spardy, L.~E., Nguyen, E.~B., Rubin, J.~E., and Berkowitz, A.
  (2011).
\newblock Strong interactions between spinal cord networks for locomotion and
  scratching.
\newblock {\em Journal of neurophysiology}, 106(4):1766--1781.

\bibitem[Horchler et~al., 2015]{horchler2015designing}
Horchler, A.~D., Daltorio, K.~A., Chiel, H.~J., and Quinn, R.~D. (2015).
\newblock Designing responsive pattern generators: stable heteroclinic channel
  cycles for modeling and control.
\newblock {\em Bioinspiration \& biomimetics}, 10(2):026001.

\bibitem[Ijspeert et~al., 2007]{ijspeert2007swimming}
Ijspeert, A.~J., Crespi, A., Ryczko, D., and Cabelguen, J.-M. (2007).
\newblock From swimming to walking with a salamander robot driven by a spinal
  cord model.
\newblock {\em science}, 315(5817):1416--1420.

\bibitem[Isidori, 1985]{isidori1985nonlinear}
Isidori, A. (1985).
\newblock {\em Nonlinear control systems}.
\newblock Springer.

\bibitem[Izhikevich, 2000]{izhikevich2000}
Izhikevich, E.~M. (2000).
\newblock Phase equations for relaxation oscillators.
\newblock {\em SIAM Journal on Applied Mathematics}, 60(5):1789--1804.

\bibitem[Izhikevich, 2007]{izhikevich2007dynamical}
Izhikevich, E.~M. (2007).
\newblock {\em Dynamical systems in neuroscience}.
\newblock MIT press.

\bibitem[Latorre et~al., 2002]{latorre2002characterization}
Latorre, R., Rodr{\'\i}guez, F.~B., and Varona, P. (2002).
\newblock Characterization of triphasic rhythms in central pattern generators
  (i): interspike interval analysis.
\newblock In {\em International Conference on Artificial Neural Networks},
  pages 160--166. Springer.

\bibitem[Lyttle et~al., 2017]{lyttle2017robustness}
Lyttle, D.~N., Gill, J.~P., Shaw, K.~M., Thomas, P.~J., and Chiel, H.~J.
  (2017).
\newblock Robustness, flexibility, and sensitivity in a multifunctional motor
  control model.
\newblock {\em Biological cybernetics}, 111(1):25--47.

\bibitem[Marder et~al., 2007]{marder2007}
Marder, E., Bucher, D., et~al. (2007).
\newblock Understanding circuit dynamics using the stomatogastric nervous
  system of lobsters and crabs.
\newblock {\em Annual review of physiology}, 69(1):291--316.

\bibitem[May and Leonard, 1975]{may1975nonlinear}
May, R.~M. and Leonard, W.~J. (1975).
\newblock Nonlinear aspects of competition between three species.
\newblock {\em SIAM journal on applied mathematics}, 29(2):243--253.

\bibitem[Molkov et~al., 2015]{molkov2015}
Molkov, Y.~I., Bacak, B.~J., Talpalar, A.~E., and Rybak, I.~A. (2015).
\newblock Mechanisms of left-right coordination in mammalian locomotor pattern
  generation circuits: a mathematical modeling view.
\newblock {\em PLoS computational biology}, 11(5):e1004270.

\bibitem[Monga and Moehlis, 2019]{monga2019optimal}
Monga, B. and Moehlis, J. (2019).
\newblock Optimal phase control of biological oscillators using augmented phase
  reduction.
\newblock {\em Biological cybernetics}, 113(1):161--178.

\bibitem[Monga et~al., 2019]{monga2019phase}
Monga, B., Wilson, D., Matchen, T., and Moehlis, J. (2019).
\newblock Phase reduction and phase-based optimal control for biological
  systems: a tutorial.
\newblock {\em Biological cybernetics}, 113(1):11--46.

\bibitem[Morrison et~al., 2016]{morrison2016diversity}
Morrison, K., Degeratu, A., Itskov, V., and Curto, C. (2016).
\newblock Diversity of emergent dynamics in competitive threshold-linear
  networks: a preliminary report.
\newblock {\em arXiv preprint arXiv:1605.04463}.

\bibitem[Olypher et~al., 2006]{olypher2006}
Olypher, A., Cymbalyuk, G., and Calabrese, R.~L. (2006).
\newblock Hybrid systems analysis of the control of burst duration by
  low-voltage-activated calcium current in leech heart interneurons.
\newblock {\em Journal of Neurophysiology}, 96(6):2857--2867.

\bibitem[Park and Rubin, 2022]{park2022}
Park, C. and Rubin, J.~E. (2022).
\newblock Activity patterns of a two-timescale neuronal ring model with
  voltage-dependent, piecewise smooth inhibitory coupling.
\newblock {\em SIAM Journal on Applied Dynamical Systems}, 21(3):1952--1999.

\bibitem[Park et~al., 2018]{park2018infinitesimal}
Park, Y., Shaw, K.~M., Chiel, H.~J., and Thomas, P.~J. (2018).
\newblock The infinitesimal phase response curves of oscillators in piecewise
  smooth dynamical systems.
\newblock {\em European Journal of Applied Mathematics}, 29(5):905--940.

\bibitem[Rubin et~al., 2011]{rubin2011}
Rubin, J.~E., Bacak, B.~J., Molkov, Y.~I., Shevtsova, N.~A., Smith, J.~C., and
  Rybak, I.~A. (2011).
\newblock Interacting oscillations in neural control of breathing: modeling and
  qualitative analysis.
\newblock {\em Journal of computational neuroscience}, 30(3):607--632.

\bibitem[Rubin et~al., 2022]{rubin2022irregular}
Rubin, J.~E., Earn, D.~J., Greenwood, P.~E., Parsons, T.~L., and Abbott, K.~C.
  (2022).
\newblock Irregular population cycles driven by environmental stochasticity and
  saddle crawlbys.
\newblock {\em Oikos}, page e09290.

\bibitem[Rubin et~al., 2009]{rubin2009multiple}
Rubin, J.~E., Shevtsova, N.~A., Ermentrout, G.~B., Smith, J.~C., and Rybak,
  I.~A. (2009).
\newblock Multiple rhythmic states in a model of the respiratory central
  pattern generator.
\newblock {\em Journal of Neurophysiology}, 101(4):2146--2165.

\bibitem[Rubin and Terman, 2002]{rubinhandbook}
Rubin, J.~E. and Terman, D. (2002).
\newblock Geometric singular perturbation analysis of neuronal dynamics.
\newblock In {\em Handbook of dynamical systems}, volume~2, pages 93--146.
  Elsevier.

\bibitem[Rubin and Terman, 2012]{rubin2012explicit}
Rubin, J.~E. and Terman, D. (2012).
\newblock Explicit maps to predict activation order in multiphase rhythms of a
  coupled cell network.
\newblock {\em The Journal of Mathematical Neuroscience}, 2(1):1--28.

\bibitem[Rybak et~al., 2006]{rybak2006modelling}
Rybak, I.~A., Shevtsova, N.~A., Lafreniere-Roula, M., and McCrea, D.~A. (2006).
\newblock Modelling spinal circuitry involved in locomotor pattern generation:
  insights from deletions during fictive locomotion.
\newblock {\em The Journal of physiology}, 577(2):617--639.

\bibitem[Sakurai and Katz, 2022]{sakurai2022bursting}
Sakurai, A. and Katz, P.~S. (2022).
\newblock Bursting emerges from the complementary roles of neurons in a
  four-cell network.
\newblock {\em Journal of Neurophysiology}, 127(4):1054--1066.

\bibitem[Schwemmer and Lewis, 2012]{schwemmer2012theory}
Schwemmer, M.~A. and Lewis, T.~J. (2012).
\newblock The theory of weakly coupled oscillators.
\newblock In {\em Phase response curves in neuroscience}, pages 3--31.
  Springer.

\bibitem[Selverston et~al., 1998]{selverston1998basic}
Selverston, A., Elson, R., Rabinovich, M., Huerta, R., and Abarbanel, H.
  (1998).
\newblock Basic principles for generating motor output in the stomatogastric
  ganglion.
\newblock {\em Annals of the New York Academy of Sciences}, 860(1):35--50.

\bibitem[Shaw, 2014]{shaw2014dynamical}
Shaw, K.~M. (2014).
\newblock {\em Dynamical Architectures for Controlling Feeding in Aplysia
  Californica}.
\newblock Case Western Reserve University.

\bibitem[Shaw et~al., 2015]{shaw2015significance}
Shaw, K.~M., Lyttle, D.~N., Gill, J.~P., Cullins, M.~J., McManus, J.~M., Lu,
  H., Thomas, P.~J., and Chiel, H.~J. (2015).
\newblock The significance of dynamical architecture for adaptive responses to
  mechanical loads during rhythmic behavior.
\newblock {\em Journal of computational neuroscience}, 38(1):25--51.

\bibitem[Shaw et~al., 2012]{shaw2012phase}
Shaw, K.~M., Park, Y.-M., Chiel, H.~J., and Thomas, P.~J. (2012).
\newblock Phase resetting in an asymptotically phaseless system: {On} the phase
  response of limit cycles verging on a heteroclinic orbit.
\newblock {\em SIAM Journal on Applied Dynamical Systems}, 11(1):350--391.

\bibitem[Skinner et~al., 1994]{skinner1994mechanisms}
Skinner, F.~K., Kopell, N., and Marder, E. (1994).
\newblock Mechanisms for oscillation and frequency control in reciprocally
  inhibitory model neural networks.
\newblock {\em Journal of computational neuroscience}, 1(1):69--87.

\bibitem[Smith et~al., 2007]{smith2007}
Smith, J.~C., Abdala, A., Koizumi, H., Rybak, I.~A., and Paton, J.~F. (2007).
\newblock Spatial and functional architecture of the mammalian brain stem
  respiratory network: a hierarchy of three oscillatory mechanisms.
\newblock {\em Journal of neurophysiology}, 98(6):3370--3387.

\bibitem[Stanhope et~al., 2014]{stanhope2014identifiability}
Stanhope, S., Rubin, J.~E., and Swigon, D. (2014).
\newblock Identifiability of linear and linear-in-parameters dynamical systems
  from a single trajectory.
\newblock {\em SIAM Journal on Applied Dynamical Systems}, 13(4):1792--1815.

\bibitem[Stone and Holmes, 1990]{stone1990random}
Stone, E. and Holmes, P. (1990).
\newblock Random perturbations of heteroclinic attractors.
\newblock {\em SIAM Journal on Applied Mathematics}, 50(3):726--743.

\bibitem[Terman et~al., 1998]{terman1998}
Terman, D., Kopell, N., and Bose, A. (1998).
\newblock Dynamics of two mutually coupled slow inhibitory neurons.
\newblock {\em Physica D: Nonlinear Phenomena}, 117(1-4):241--275.

\bibitem[Wang and Rinzel, 1992]{wang1992alternating}
Wang, X.-J. and Rinzel, J. (1992).
\newblock Alternating and synchronous rhythms in reciprocally inhibitory model
  neurons.
\newblock {\em Neural computation}, 4(1):84--97.

\bibitem[Wang et~al., 2021]{wang2021shape}
Wang, Y., Gill, J.~P., Chiel, H.~J., and Thomas, P.~J. (2021).
\newblock Shape versus timing: linear responses of a limit cycle with hard
  boundaries under instantaneous and static perturbation.
\newblock {\em SIAM Journal on Applied Dynamical Systems}, 20(2):701--744.

\bibitem[Wang et~al., 2022]{Wang2022variational}
Wang, Y., Gill, J.~P., Chiel, H.~J., and Thomas, P.~J. (2022).
\newblock Variational and phase response analysis for limit cycles with hard
  boundaries, with applications to neuromechanical control problems.
\newblock {\em Biological cybernetics}, 116(2):687–710.

\bibitem[Webster-Wood et~al., 2020]{webster2020control}
Webster-Wood, V.~A., Gill, J.~P., Thomas, P.~J., and Chiel, H.~J. (2020).
\newblock Control for multifunctionality: bioinspired control based on feeding
  in {\textit{{{aplysia}} californica}}.
\newblock {\em Biological cybernetics}, 114(6):557--588.

\bibitem[Weimann et~al., 1991]{weimann1991neurons}
Weimann, J.~M., Meyrand, P., and Marder, E. (1991).
\newblock Neurons that form multiple pattern generators: identification and
  multiple activity patterns of gastric/pyloric neurons in the crab
  stomatogastric system.
\newblock {\em Journal of neurophysiology}, 65(1):111--122.

\bibitem[Whalen et~al., 2015]{whalen2015observability}
Whalen, A.~J., Brennan, S.~N., Sauer, T.~D., and Schiff, S.~J. (2015).
\newblock Observability and controllability of nonlinear networks: The role of
  symmetry.
\newblock {\em Physical Review X}, 5(1):011005.

\bibitem[Wiener, 1948]{wiener1948cybernetics}
Wiener, N. (1948).
\newblock {\em Cybernetics or Control and Communication in the Animal and the
  Machine}.
\newblock MIT press.

\bibitem[Wilson and Moehlis, 2022]{wilson2022recent}
Wilson, D. and Moehlis, J. (2022).
\newblock Recent advances in the analysis and control of large populations of
  neural oscillators.
\newblock {\em Annual Reviews in Control}.

\bibitem[Wojcik et~al., 2014]{wojcik2014}
Wojcik, J., Schwabedal, J., Clewley, R., and Shilnikov, A.~L. (2014).
\newblock Key bifurcations of bursting polyrhythms in 3-cell central pattern
  generators.
\newblock {\em PloS one}, 9(4):e92918.

\bibitem[Yakovenko et~al., 2005]{yakovenko2005}
Yakovenko, S., McCrea, D., Stecina, K., and Prochazka, A. (2005).
\newblock Control of locomotor cycle durations.
\newblock {\em Journal of Neurophysiology}, 94(2):1057--1065.

\bibitem[Yu and Thomas, 2021]{yu2021dynamical}
Yu, Z. and Thomas, P.~J. (2021).
\newblock Dynamical consequences of sensory feedback in a half-center
  oscillator coupled to a simple motor system.
\newblock {\em Biological cybernetics}, 115(2):135--160.

\bibitem[Zhang and Lewis, 2013]{zhang2013phase}
Zhang, C. and Lewis, T.~J. (2013).
\newblock Phase response properties of half-center oscillators.
\newblock {\em Journal of computational neuroscience}, 35(1):55--74.

\end{thebibliography}
\bibliographystyle{apalike}
\end{document}